\documentclass{elsart}
\usepackage{graphics}
\usepackage{epsfig}
\begin{document}
\begin{frontmatter}
\title{Distributions of secondary muons at sea level from cosmic
       gamma rays below 10~TeV}
\author[Poirier]{J.\ Poirier,}
\author[Roesler]{S.\ Roesler,}
\author[Fasso]{A.\ Fass\`o}

\address[Poirier]{Center for Astrophysics, Physics Department, University 
                  of Notre Dame, Notre Dame, Indiana 46556, USA}
\address[Roesler]{Stanford Linear Accelerator Center, MS. 48, 
                  2575 Sand Hill Road,\\ Menlo Park, California 94025, USA\\
                  Tel.: +1-650-926-2048, Fax: +1-650-926-3569\\ 
                  Email: sroesler@slac.stanford.edu}
\address[Fasso]{CERN-EP/AIP, CH-1211 Geneva 23, Switzerland}
%
%
\begin{abstract}
The FLUKA Monte Carlo program is used to predict the distributions of
the muons which originate from primary cosmic gamma rays and reach sea level.
The main result is the angular distribution of muons produced by vertical
gamma rays which is necessary to predict the inherent angular resolution of 
any instrument utilizing muons to infer properties of gamma ray primaries.
Furthermore, various physical effects are discussed which affect these 
distributions in differing proportions.
\end{abstract}
\end{frontmatter}
%
%
\section{Introduction}
\label{intro.sec}
Muons detected at ground level arise mainly as decay products of
charged mesons.  These mesons are created abundantly in hadronic showers,
most of which are caused by primary cosmic ray protons
and nuclei interacting inelastically with the nuclei of the
atmosphere. A small fraction of muons, however, have their first
origin in photonuclear reactions induced by primary 
or secondary cosmic gamma rays.
Several authors have performed analytical or Monte Carlo calculations 
of the muon flux produced in gamma showers (see references [10-21] of 
\cite{Fas00c}). 
Most of these calculations were one dimensional and all of them 
referred to gamma energies much larger than 10 TeV.  

Despite their relative scarcity compared to the large background of
muons from hadron-generated showers, muons originating from primary gamma
ray interactions are important for ground-based high statistics cosmic ray
experiments which are sensitive to energies $\le 10$~TeV such as 
MILAGRO~\cite{MILAGRO} and GRAND~\cite{GRAND}. Cosmic gamma rays, unlike 
charged hadrons, are unaffected by the Earth's and galactic magnetic fields 
and their direction points directly to the location of their source. 
Therefore, an
accumulation of muon directions around a particular angle (and also at a  
particular time in the case of pulsed sources or gamma ray bursts)
carries direct information about the location of the source and the 
neutrality of the primary causing the excess.  
The ability to distinguish these muons from background muons depends 
upon statistics, the strength and energy spectrum of the emitting source, 
the angular resolution of the experiment, and 
the degradation of the angular resolution due to all the physical processes 
which occur between the primary and the muon at detection level.  
This last point is the focus of this calculation.  
Knowledge of this angular resolution is important in order to
determine a detection window which is large enough to include a
good fraction of the gamma
signal and, at the same time, small enough to minimize
the large uniform background of hadronic origin.

The information about the primary gamma direction is degraded 
by multiple physical processes so the final measured direction of the 
muon is no longer collinear with the primary which originated 
the muon's ancestors.
In particular, the direction resolution is affected by the angle of production 
of the mesons 
in the primary interaction (and then in subsequent hadronic interactions 
of these mesons to create additional mesons), 
the decay angle of the muon relative to its parent meson, and 
the lengths of the charged particle paths including that of the 
detected muon at ground level.  
Along these path lengths there is
Coulomb scattering and deflection in the Earth's magnetic field for 
the trajectories of the mesons and the muon (usually 
most important for the muon which typically has the longest 
flight path and the lowest momentum.) 
In addition, these effects depend upon the
energy of the particles involved and on the varying properties of 
the atmosphere (pressure, temperature.)
The muons are produced by mesons of
second, third, or further generations.
The number of generations
typically increases with increasing primary gamma energy and each
new generation involves an additional production angle which tends
to increase the angular disparity between muon and
primary because of more steps in a random walk type process.  
An opposing effect 
at higher energies is that all angular distributions are more
forward-peaked thus decreasing the angular disparity;
the resulting effect is a combination of the two.

The purpose of the present study is to estimate the influence of the
various effects on the ultimate angular resolution which can be obtained 
utilizing ground level muons. The complexity
of the problem requires detailed analyses which can only be obtained
by a full three-dimensional Monte Carlo (MC) calculation. In addition,
since only a small fraction of all primary photons gives rise to an
event of interest, 
it is desirable to resort to statistical variance
reduction techniques to bring the computing task within manageable limits.

The Monte Carlo program {\sc FLUKA}~\cite{Fas97a,Fas97b,Fas00a,Fas00b} 
provides an accurate
and well-tested description of the hadronic and electromagnetic 
interactions with 
all the relevant physical effects, and
a set of effective statistical tools to accelerate convergence of the
results.  
The program has been used to calculate the geometrical
properties of secondary muons at sea level arising from primary gamma
cosmic rays with energies from 1~GeV to 10~TeV interacting in
the Earth's atmosphere. 
In this energy range the cross section for
direct muon pair production is much smaller than that for
hadroproduction and has been neglected. 
This range has been shown in~\cite{Fas00c} to be the interval which 
produces the maximum secondary muon flux at ground level for primary
differential gamma spectra with spectral indices larger than 2.4 
(a smaller index would yield a harder spectrum and give a maximum 
secondary muon 
flux arising from primary energies beyond those considered here.)

In this paper, only incident cosmic gamma rays at perpendicular (vertical) 
incidence on top of the atmosphere are considered.  
Ground-based experiments often choose angles close 
to normal for reasons of simplicity and because the detection 
rate is a maximum, the rate dropping rather rapidly as the 
angle deviates markedly from normal.  For small angles from normal, 
the angular resolution results will probably 
not differ significantly from those presented in this paper.

In addition to the angular correlation of secondary muons at 
ground level, this study provides also additional information 
such as the distribution of heights at
which the muons originate, the muon's kinetic energy distribution at the
ground and production level, 
the radial distribution of the muons, and the number of generations which 
preceded that of the muon.
%
%
\section{Air shower calculations with {\sc FLUKA}}
\label{fluka.sec}
The 2000 version of the {\sc FLUKA} Monte Carlo code
was used to simulate the electromagnetic and hadronic particle cascades induced
by primary gamma rays in the atmosphere. The powerful biasing capabilities
of the code allowed a simulation of the full three-dimensional shower in a 
single run from the top of the atmosphere down to ground level. 

The physical
models implemented in {\sc FLUKA} which are relevant for the present study 
as well as various variance reduction techniques were already discussed
in~\cite{Fas00c}. There it was emphasized that these models were validated
by comprehensive comparisons of {\sc FLUKA} results with experimental data
obtained mostly at accelerators but also in recent cosmic ray 
studies \cite{Pat95,Fer97a,Roe98a,Bat00}.
Nevertheless, the simulation of an atmospheric shower and, in particular, of
its muon component is a relatively new area of application of {\sc FLUKA}. 
Additional examples showing 
that the code allows remarkably accurate predictions in this area 
are given in Section~\ref{fluka-datcmp.subsec}.
Section~\ref{fluka-simdet.subsec} contains some details of the present 
gamma ray shower 
simulations which supplement those discussed in~\cite{Fas00c}; 
this reference contains a more complete description.   
%
\subsection{Muon production in air showers induced by hadronic primaries 
            and comparison to experimental data}
\label{fluka-datcmp.subsec}
No data exist for the muon component of gamma ray-induced 
showers to which {\sc FLUKA} results could be directly compared. 
On the other hand, data on muon 
flux and energy spectra have become available recently for atmospheric 
showers induced by cosmic ray protons and nuclei. These data were mostly 
obtained with spectrometers in balloon-borne experiments and cover the whole 
range of atmospheric depths from the top of the atmosphere down to sea level.

Fortunately, the mechanisms and models for muon production and transport in
gamma ray-induced showers are largely the same as those in pure hadronic
air showers. In both cases, the production of the pions which eventually
decay into muons is described by the Dual Parton Model (DPM)~\cite{Cap94}. The
only difference is that in the former case the photon is assumed to fluctuate
first into a quark-antiquark state which then interacts hadronically, identical
to a meson. This so-called Vector Meson Dominance model (VMD) for the hadronic
photon fluctuation is a well-established concept (see~\cite{Bau78} 
and references therein) and its application together with the DPM to
hadronic interactions of photons has been proven to be very successful 
(see, for example, \cite{Eng95,Roe98b}). A good description of muon 
production from proton- and nuclei-induced air showers
should therefore give confidence that the corresponding results 
obtained with {\sc FLUKA} for photon-induced air showers are reliable as well.
In the following a few examples are given.

A critical aspect of muon production in the atmosphere is the dependence
of the total muon flux on the height above sea level. 
This dependence is commonly
expressed as a function of depth in the atmosphere (in g/cm$^2$ of air) 
which is a measure of the amount of air penetrated by the cascade. It
has been measured by the CAPRICE experiment which was able to
distinguish between positively and negatively charged muons. Results obtained
with {\sc FLUKA} for negative muons are compared with CAPRICE data~\cite{Boe00a}
in Fig.~\ref{mu-capdep}.
The different curves correspond to the muon flux calculated for different 
intervals of the muon momentum. To correspond to the average geometrical 
response of the CAPRICE apparatus, only muons with a polar angle of less 
than $9^\circ$ with respect to the vertical (i.e., vertical muons) were scored. 
The simulations are based on sophisticated models and methods for the 
sampling of the identity, energy, and angle of the hadronic primary from
the distributions of galactic cosmic rays on top of
the atmosphere at the time of the measurements. More details can be found 
in~\cite{Roe98a,Roe00}. As can be seen from the figure, {\sc FLUKA} gives
a good description of the negative muon flux for all depths. 
Similar agreement exists for positive muons (not shown here).

Energy spectra of muons were recorded by the same experiment
at ground (i.e., at an atmospheric depth of 
886~g/cm$^2$)~\cite{Kre99}. {\sc FLUKA} results for positive and negative muons
are compared to these data in Fig.~\ref{mu-capspc25&ratio}a. 
Again, good agreement between simulation and data is found.

Finally, the ratio of the fluxes of positively and negatively charged muons
is compared to CAPRICE data~\cite{Kre99} in Fig.~\ref{mu-capspc25&ratio}b. 
Data and calculations are again for vertical muons at a depth of 886~g/cm$^2$. 
The ratios are plotted as a function of the muon energy.

The above comparisons of {\sc FLUKA} results and experimental data together 
with those published in earlier studies~\cite{Pat95,Fer97a,Roe98a,Bat00}
indicate that the application of {\sc FLUKA} to cosmic gamma ray showers
should give reliable predictions as well.
%
\subsection{The Monte Carlo simulation of cosmic gamma ray showers}
\label{fluka-simdet.subsec}
In order to study the dependence of the shower properties on the primary
photon energy, the showers were calculated for monoenergetic photons
impinging vertically on top of the atmosphere (taken to be 
80~km above sea level). 
In total 9 sets of simulations were performed for the following primary 
energies: 1, 3, 10, 30, 100, 300, 1000, 3000, and 10000~GeV. As in the earlier 
study~\cite{Fas00c} the atmosphere was approximated by 50 layers with 
a constant density in each layer and with 
layer-densities decreasing exponentially 
with increasing altitude. It has been verified~\cite{Fas00c}
that this approximation does not affect the conclusions drawn in this study.
The geometry is described in a right-handed orthogonal system with its
origin at the intersection of the shower axis with sea-level elevation,
$z$ points to the center of the
Earth and $x$ and $y$ point North and East, respectively. Most of the
results discussed throughout the paper were obtained without the 
effect of the magnetic field of the Earth included.  
However, for two primary energies (10~GeV and 1~TeV) 
additional simulations were performed which included the magnetic field of 
three different geographical locations (see Sec.~\ref{angular-magfld.subsec}).

The use of several variance reduction (biasing) techniques was essential to 
obtain results with reasonable statistical significance. They included
leading particle biasing at each electromagnetic interaction, biasing of
the photon mean-free-path with respect to photonuclear interactions,
biasing of the decay length of charged mesons, and particle splitting
at the boundaries of different air layers. Each of these biasing techniques
alters the statistical weight of a particle in the cascade.
A muon as produced and transported in the Monte Carlo simulation and which
carries a certain statistical weight (hereafter called ``MC muon'') contributes
to distributions, yields, etc., of actual, i.e., measurable muons with a
probability equal to its weight. The use of biasing techniques would not be
appropriate to study fluctuations within the same
shower which, however, is not the aim of this study.
More details on the biasing techniques can be found in~\cite{Fas00c}.

For each MC muon reaching sea level (detection level) the following information
was recorded in a file for later analysis:
\begin{enumerate}
 \item Information on the muon at detection level: 
       muon charge,
       lateral coordinates $(x,y)$ with respect to shower axis, 
       direction cosines with respect to the $x$ and $y$ axes,
       kinetic energy, statistical weight of the muon, and
       number of the event (i.e., primary cosmic ray photon) which
       produced this muon.
 \item Information on the muon at its production vertex (i.e., meson decay
       vertex):
       lateral coordinates $(x,y)$ with respect to shower axis,
       direction cosines with respect to the $x$ and $y$ axes,
       height $z$,
       kinetic energy, statistical weight of the muon, and
       identity of the decaying meson.
 \item Information on the grandparent at the parent production vertex:
       particle identity,
       lateral coordinates $(x,y)$ with respect to shower axis,
       direction cosines with respect to the $x$ and $y$ axes,
       height $z$,
       kinetic energy, statistical weight of the particle, and
       generation number.
 \item Information on the photonuclear interaction vertex preceding
       the hadronic cascade in which the muon at detection level has been
       created:
       lateral coordinates $(x,y)$ with respect to shower axis,
       direction cosines with respect to the $x$ and $y$ axes,
       height $z$,
       energy, statistical weight of the photon, and
       generation number of the photon
\end{enumerate}
The generation of a particle increases with each sampled discrete
interaction (electromagnetic or hadronic), i.e., the primary photon
is generation ``1,'' the generation of the electron and positron after
the first pair production process would be ``2,'' etc. Delta ray production
and Coulomb scattering are not considered to increase the generation number.

A summary of the simulated particles cascades for each primary gamma ray
energy is given in Table~\ref{no-of-muons.tab}. In the following,
all results refer to the sum of positive and negative muons. 
The number of histories
(number of primary gamma rays), $N_{\gamma}$, calculated for each primary 
energy decreases with energy in order to obtain uniform statistical
significance on the muons which reach ground level 
for the different primary energies. At the lowest energy (1~GeV) the muon 
flux per gamma at sea level is very small so that even
with a significant amount of biasing in the simulations it is difficult to
reach adequate statistics. 
Less emphasis was therefore put on the simulation at 1~GeV.
The column $N_{\mu}$ gives the total number of muons reaching detection
level for the given number of primary photons $N_\gamma$.
The predictions for the muon multiplicity at sea level per incident primary 
gamma ray $N_{\mu}/N_{\gamma}$ are listed in the last column.
Similar values were already reported in~\cite{Fas00c}, however with 
smaller statistical significance than those in Table~\ref{no-of-muons.tab}.
\begin{table}[htbp]
 \caption{\label{no-of-muons.tab}
    Summary of the simulated particle cascades. $N_{\gamma}$ is the number 
    of histories calculated for each primary energy ($E_{\gamma}$) and $N_{\mu}$
    is the total number of muons (i.e., the sum of the weights) scored at 
    sea level. In the last column
    the average muon multiplicity per primary gamma ray is given.
    The errors quoted in the last column represent the statistical uncertainties
    of the calculations.
         }
 \vspace{2mm}
 \begin{center}
  \begin{tabular} {|c||c|c|c|}
    \hline 
    $E_{\gamma}$ (GeV) & $N_{\gamma}$ & $N_{\mu}$ & $N_{\mu}/N_{\gamma}$ \\
    [0.5ex] \hline \hline
    1    & 4542000  &  $3.78\times 10^{-3}$ & $(8.32\pm 0.46)\times 10^{-10}$ \\
    3    &26000000  &  117.7 & $(4.53\pm 0.23)\times 10^{-6}$  \\
    10   & 2250000  &  1191  & $(5.29\pm 0.03)\times 10^{-4}$  \\
    30   & 1260000  &  3843  & $(3.05\pm 0.01)\times 10^{-3}$  \\
    100  &  800000  &  11351 & $(1.42\pm 0.01)\times 10^{-2}$  \\
    300  &  324445  &  17259 & $(5.32\pm 0.02)\times 10^{-2}$  \\
    1000 &  143131  &  31709 & $(2.22\pm 0.01)\times 10^{-1}$  \\
    3000 &   27040  &  21954 & $(8.12\pm 0.05)\times 10^{-1}$  \\
    10000&   18000  &  58874 & $3.27\pm 0.03 $               \\
  [0.5ex] \hline
  \end{tabular}
  \end{center}
\end{table}
%
%
\section{Angular correlations}
\label{angular.sec}
%
\subsection{Angular distribution of muons at sea level}
\label{angular-dissl.subsec}
In this section the effect of the Earth's magnetic field is not included; 
Section~\ref{angular-magfld.subsec} has a separate discussion of this 
added effect which depends 
on the geographical location of interest.
Fig.~\ref{dNdTh} summarizes the deviations of the angle of the muons which
reach sea level from the primary gamma direction (vertical angle).
It shows the distribution of the angle $\Theta_{xz}$ which is the 
muon's angle with respect to the $z$ axis 
projected onto the $xz$ plane (North-down plane).
This angle is defined as $\Theta_{xz}=\arctan 
(\cos \theta_x/\cos \theta_z)$ where $\cos \theta_x$ and $\cos \theta_z$
are the direction cosines of the detected muon. Negative projections have 
been reflected upon the positive values for $\Theta_{xz}$ since positive 
and negative angles would be symmetric in the absence of
the magnetic field. Also, the distributions in the $yz$ plane 
(East-down plane) are similar to those shown in Fig.~\ref{dNdTh}.

Since the angle of the primary gamma with respect to the $z$ axis is 
zero (vertically incident primaries), any deviation of the muon direction from 
zero degrees is due to the various interaction and transport processes 
occurring in the cascade between the primary gamma ray and the muon which 
reaches sea level. Thus the results shown in Fig.~\ref{dNdTh} represent the 
correlation between the direction of the detected muon and the direction 
of the primary gamma ray.

All distributions in Fig.~\ref{dNdTh} are normalized to unit area to compare 
their shapes.
The shape of the distributions for primary energies above about 30~GeV
varies rather slowly. Therefore, only two histograms
(100~GeV and 10~TeV) are shown for this energy range. At lower energies
the shape of the distributions changes significantly as can be seen from
the histograms for 3 and 10~GeV. The distributions become wider with
decreasing primary energy. At low primary gamma ray energies, 
the muons have, on average, lower energies which cause larger angular 
deviations from the primary due to kinematics and scattering effects
which are larger for the lower energies. This
is illustrated in Fig.~\ref{dNdTh-10GTeV} for primary energies of 10~GeV and 
10~TeV. An arbitrary cut of 2~GeV was applied to the angular distributions
so that contributions from muons with energies below and above this cut
could be seen separately along with the total.  

Two measures of the widths of these distributions as a function of the
energy of the incident primary cosmic gamma ray are given in 
Table~\ref{angwidths.tab}. The width parameters are: the half-width at 
\begin{table}[htbp]
 \caption{\label{angwidths.tab}
    Half-width at half-maximum-height ($\delta\Theta_{xz}$(HWHM)) of the 
    angular distributions and angular half-width containing 68\% of the muons
    ($\delta\Theta_{xz}$(68\%)).
    For both quantities the values are given for all muon energies and
    for muon kinetic energies above 1, 2, and 4~GeV. Units of the widths are 
    in degrees.
         }
 \vspace{2mm}
 \begin{center}
  \begin{tabular} {|c||c|c|c|c|}
    \hline
    & \multicolumn{4}{c|}{$\delta\Theta_{xz}$(HWHM)} \\
    [0.5ex] \hline
    $E_{\gamma}$ (GeV) 
    & $E_\mu>0$ & $E_\mu>1$ GeV & $E_\mu>2$ GeV & $E_\mu>4$ GeV \\
    [0.5ex] \hline \hline
    1    & 31.0  & $-$  & $-$  & $-$  \\
    3    & 8.20  & $-$  & $-$  & $-$  \\
    10   & 3.23  & 2.95 & 2.57 & 1.87 \\
    30   & 1.87  & 1.73 & 1.59 & 1.35 \\
    100  & 1.12  & 1.04 & 0.95 & 0.81 \\
    300  & 0.92  & 0.85 & 0.77 & 0.66 \\
    1000 & 0.91  & 0.82 & 0.74 & 0.63 \\
    3000 & 0.98  & 0.89 & 0.78 & 0.66 \\
    10000& 1.05  & 0.94 & 0.82 & 0.67 \\
    [0.5ex] \hline
    & \multicolumn{4}{c|}{$\delta\Theta_{xz}$(68\%)}\\
    [0.5ex] \hline
    1    & 28.3  & $-$  & $-$  & $-$  \\
    3    & 7.36  & $-$  & $-$  & $-$  \\
    10   & 3.83  & 3.06 & 2.46 & 1.66 \\
    30   & 3.07  & 2.40 & 1.95 & 1.45 \\
    100  & 2.87  & 2.15 & 1.69 & 1.21 \\
    300  & 3.04  & 2.18 & 1.68 & 1.16 \\
    1000 & 3.32  & 2.31 & 1.75 & 1.19 \\
    3000 & 3.62  & 2.42 & 1.81 & 1.23 \\
    10000& 4.05  & 2.57 & 1.89 & 1.27 \\
  [0.5ex] \hline
  \end{tabular}
  \end{center}
\end{table}
half-maximum-height (HWHM) and the angular half-width containing 68\% of the 
muons, i.e., the same number of events that a 1$\sigma$ cut would include if
the distributions had a Gaussian shape
(which they don't have, because of the long tails). 
Both of these parameters provide a measure of the distribution widths which
minimizes the distorting effect of a tail.
In addition, the values for various cuts on the kinetic 
energy of the detected muons (1, 2, and 4~GeV) are presented. The cutoff 
energy results for primary energies of less than 10~GeV were omitted due to 
lack of statistics. For the HWHM calculation, 
the maximum height is assumed to be the value
in the first bin of the corresponding histogram, i.e., the average value
in the interval $|\Theta_{xz}| < 0.25$ degrees.
The $x$ and $y$ rectangular coordinate system was chosen to represent the 
results of the muon's angular distribution as the corrections for the 
additional deflection due to the Earth's magnetic field differ in these 
two directions. In addition, $(x,y)$ is the natural coordinate system for 
GRAND. For some experiments a space angle is more natural.  If we define 
$R = \sqrt{x^2+y^2}$, then the space angle resolution 
($\delta\Theta_{R}$(HWHM)) can be estimated directly 
from the numbers in Table~\ref{angwidths.tab} by multiplying the 
$\delta\Theta_{xz}$(HWHM) values 
by $\sqrt{2}$ due to the symmetry of $x$ and $y$ in the absence of 
a magnetic field.  

As can be seen in this table: (i) as the primary energy decreases below 10~GeV, 
the width of the angular distribution increases dramatically,
(ii) as the energy rises above 300~GeV, there is only a small, gradual 
increase in the width of the angular distribution, and (iii) the width
of the distribution narrows as the muon's cutoff energy is raised; 
i.e., the angular resolution becomes better 
by eliminating the lower energy muons which have, on average, poorer 
angular resolution. 
%
\subsection {Correlation of muon angle with distance from the shower axis}
\label{angular-corpos.subsec}
In the preceding section, angular distributions were given for all muons
regardless of their lateral distance from the shower axis. However, as the 
distance from the shower axis increases, the angle is systematically 
biased away from the shower axis.  
Fig.~\ref{dNdThx-10GeV1TeVx} shows the correlation between 
the projected angle $\Theta_{xz}$ and the $x$ coordinate (i.e., the
northward distance from the shower core) of the muon position 
at sea level for primary gamma ray energies of 10~GeV and 1~TeV.
The angular distributions are given for different intervals in $x$ and
each distribution is normalized to unit area.
Here, negative values of $\Theta_{xz}$ have not been reflected onto 
positive values and are explicitly shown.  
As expected, the peak of the $\Theta_{xz}$ distributions 
are shifted toward positive values as $x$ becomes more positive 
and the distributions become wider as $x$ 
increases. Both effects depend on the energy of the primary gamma ray and
are more pronounced at higher energies.

Hence, the angular distributions narrow significantly if they are limited
to muons within a certain interval around the shower axis.
Information on this correlation between the angle and the lateral
distance of the muon from the shower axis allows experiments to improve
the angular resolution by applying cuts to the data, e.g., in $x$.  
If the available statistics demand using all available muons, their 
angle at large $x$'s ($y$'s) could be corrected by the most probable angle at 
that $x$ (or $y$) to obtain a better estimate of the primary's direction.  

This correlation can also be used to infer the height from 
which the muon originated, a parameter difficult to obtain 
experimentally. If one extrapolates the muon direction backward (upward), 
the intersection point with the centroid of the shower 
(which can be obtained from the more numerous electrons in 
the same shower), 
is an estimator for the height where the muon was created.
However, deviations of the muon production vertex from the centroid and 
the physical processes which alter the muon track between the production 
vertex and sea level obscure this information somewhat.  
For this reason it is not 
possible to study this question on a muon-by-muon basis. For 
example, tracks with projected angles close to zero which exist at almost 
all $x$-distances from the origin would yield, upon extrapolation, an 
infinite height-of-origin. This problem can be circumvented by 
finding the mean $x$ within an interval of $x$ and dividing this mean 
by the corresponding mean of $\tan\Theta_{xz}$.  
The averaging serves another purpose as well: the randomness 
of the various physical processes, such as scattering, tend 
to cancel in the averaging.  
%

\subsection {Angular resolution for a spectrum of primary gamma
             ray energies}
\label{angular-spec.subsec}
The results of the preceding sections can be combined to obtain the expected 
angular resolution for a spectrum of primary gamma rays. Various steps in 
the calculation are shown in Fig.~\ref{flxgamma}a. Here, a differential 
energy spectrum $d\Phi_\gamma/dE_\gamma \propto E_\gamma^{-\alpha}$ with a spectral
index of $\alpha=2.41$ is assumed; this $\alpha$ corresponds to the average 
of the spectral indices reported in the Third EGRET Catalog~\cite{Har99}.
Folding $d\Phi_\gamma/dE_\gamma$  with the number of muons reaching sea level 
per primary photon, $N_\mu/N_\gamma$, yields the number of muons at sea 
level as function of the primary gamma ray energy, $d\Phi_\mu/dE_\gamma$, with
$\Phi_\mu=N_\mu/N_\gamma \times\Phi_\gamma$. Note that the differential spectra
in Fig.~\ref{flxgamma}a are multiplied by $E_\gamma$.  

The flatness of the differential muon flux per gamma at sea level 
($E_\gamma\times d\Phi_\mu/dE_\gamma$) above 10~GeV signifies that the muons 
originate rather uniformly from a broad range of primary energies.  
Below 10~GeV, the muon flux decreases steeply as $N_\mu/N_\gamma$ rapidly 
approaches zero. 
Harder spectral indices ($\alpha < 2.41$) would enhance the muon 
flux from higher primary energies and, conversely, softer indices would
enhance lower energies.

Furthermore, in Fig.~\ref{flxgamma}a the half-widths at half-maximum-height 
$\delta\Theta_{xz}$ (HWHM) 
(given in Table~\ref{angwidths.tab} in the ``$E_\mu>0$'' column) are 
plotted.
These widths are multiplied by the differential muon flux which is also
shown in Fig.~\ref{flxgamma}a. As can be seen, the larger angular 
widths at the 
lower energies ($E_\gamma<3$~GeV) do not contribute due to the small sea level 
muon flux at these energies. 

The average angular resolution (i.e., average angular width) in the North-down 
plane, $\langle \delta\Theta_{xz}{\rm (HWHM)}\rangle$, for a differential 
gamma ray spectrum with an index $\alpha$ is obtained from
\begin{equation}
\langle \delta\Theta_{xz}{\rm (HWHM)}\rangle=\frac
  {\int \delta\Theta_{xz}\times N_\mu/N_\gamma \times E_\gamma^{-\alpha}\ dE_\gamma}
  {\int N_\mu/N_\gamma \times E_\gamma^{-\alpha}\ dE_\gamma}
\end{equation}
and shown in Fig.~\ref{flxgamma}b. As can be seen, the resolution is
about constant ($1^\circ$) up to a spectral index of 2 and then rises for
larger indices due to the increasing importance of the wider angular 
widths at low energy.
%
\subsection{Effect of the magnetic field}
\label{angular-magfld.subsec}
The magnetic field of the Earth deflects positively (negatively) 
charged particles 
primarily eastward (westward) causing the angular distributions in the
projected angle $\Theta_{yz}$ (i.e., projected onto the East-down plane) 
to become wider. The size of the effect
depends on the strength and direction of the field and, therefore, 
varies with geographic location.
To obtain a measure of the magnitude of the added angular widths 
and how it depends on geographic location, three different
locations are considered and summarized in Table~\ref{magfld.tab}. 
In order to estimate the maximum possible
deflection, a location of maximum northward field component was
chosen. Due to the offset of the magnetic field axis from
the Earth's center, this location is at about $9^\circ$N, $100^\circ$E
(below referred to as ``maximum''). In addition, two locations 
with intermediate magnetic fields were chosen:  
$42^\circ$N, $86^\circ$W (GRAND) 
and $36^\circ$N, $106^\circ$W (MILAGRO). The magnetic field components
for these locations 
were obtained from the latest revision of the International 
Geomagnetic Reference Field (IGRF)~\cite{IGRF2000} and are listed in 
Table~\ref{magfld.tab}.  Here, the components of the magnetic field, B, 
are given in nanotesla (nT) and the components $B_x, B_y, B_z$ are directed
North, East, and toward the center of the Earth, respectively. 
For uniformity of comparison, all values are for sea level elevation.
\begin{table}[htbp]
 \caption{\label{magfld.tab}
    Earth's magnetic field components $B_{x}$ (North), 
    $B_{y}$ (East), and $B_{z}$ (vertically downward) 
    in nanotesla for the three considered locations. 
    The field has been
    obtained from the IGRF2000 model~\protect\cite{IGRF2000}. 
    For a given location the field components vary only by $\le 4$\% 
    for heights of 0 to 80 km.  Here,
    the values are given for sea level and a date in Feb., 2001.
         }
 \vspace{2mm}
  \begin{center}
  \begin{tabular} {|c||c|c|c|}
    \hline
    Location & $B_{x}$ (nT) & $B_{y}$ (nT) & $B_{z}$ (nT) \\
    [0.5ex] \hline \hline
    (1) GRAND, $42^\circ$N, $86^\circ$W    & 18592 &  -1361 & 52547 \\
    [0.5ex] \hline
    (2) MILAGRO, $36^\circ$N, $106^\circ$W & 22751 &   4141 & 46241 \\
    [0.5ex] \hline
    (3) Maximum, $9^\circ$N, $100^\circ$E  & 41407 &   -211 &  1326 \\
  [0.5ex] \hline
  \end{tabular}
  \end{center}
\end{table}

{\sc FLUKA} is able to transport particles in arbitrary magnetic fields.
However, since the variation of the field in the atmosphere with height above 
sea level and with time is only minor (about 4\%), for simplicity 
the magnetic field used in the calculation for a
certain location was assumed to be constant throughout the whole geometry.
In addition, although the
MILAGRO experiment is located at a height of 2170~m above sea level, all
results reported for that location refer to sea level as this allows a 
direct comparison to the results at the other locations.  

The angular distributions at sea level for the three magnetic field conditions
and for primary gamma ray energies of 10~GeV and 1~TeV are shown in 
Fig.~\ref{dNdTh-10GeV1TeVmag}.
In addition, the distributions obtained without magnetic field are given.
As can be seen, for a 10~GeV gamma primary, the effect of 
the Earth's magnetic field is considerable and changes with geographic 
location; as the northward component of the magnetic field increases, 
the width increases.  However, at 1~TeV, there is only a 
slight widening of the distributions with increasing the North component of the
field.  To within the statistical uncertainties of the calculations, 
the total number of muons which reach sea level per primary gamma ray 
is independent of the effect of the Earth's magnetic field for the two energies 
studied in this section (10~GeV and 1~TeV).  

The half-widths of the angular distributions in the North-down ($\Theta_{xz}$) and 
East-down planes ($\Theta_{yz}$) containing 68\% of the muons are presented
for the different locations in Table~\ref{angwidthsmag.tab}. Again, the
\begin{table}[htbp]
 \caption{\label{angwidthsmag.tab}
    The half-widths of the muon's sea level angular distributions which contain 
    68\% of the muons and their dependence on the magnetic field strength.  
    Results are presented for three different geographic locations: 
    (1) $42^\circ$N, $86^\circ$W (GRAND), (2) $36^\circ$N, $106^\circ$W 
    (MILAGRO), and (3) $9^\circ$N, $100^\circ$E (maximum) in order of 
    increasing values of $B_{x}$. For each location the widths are given
    for the projected angles onto the 
    North-down (labeled ``N-d.'') and East-down (labeled ``E-d.'')
    planes.
    The reference values from the calculations with no magnetic field 
    (labeled ``n.f.'') are also shown; here, 
    the small differences between North-down and East-down are only  
    statistical 
    fluctuations. Widths are listed for muon kinetic energies
    above 0 (all), 1, 2, and 4~GeV for primary gamma energies of 
    10 and 1000~GeV. The additional column for $E_\mu>0$ gives
    the widths in the East-down plane of the muons 
    at their production vertex (``birth'', labeled ``b'').
    Units of the widths are in degrees.
         }
 \vspace{2mm}
  \begin{center}
  \begin{tabular} {|c|c||c|c|c||c|c||c|c||c|c|}
    \hline
    & $E_{\gamma}$ 
    & \multicolumn{3}{c||}{$E_\mu>0$} & \multicolumn{2}{c||}{$E_\mu>1$ GeV} 
    & \multicolumn{2}{c||}{$E_\mu>2$ GeV}& \multicolumn{2}{c|}{$E_\mu>4$ GeV}\\
    [0.5ex] \hline
    & (GeV) & N-d. & E-d. & E-d.(b) 
            & N-d. & E-d. & N-d. & E-d. & N-d. & E-d. \\
    [0.5ex] \hline \hline
    n.f.
        &  10& 3.83 & 3.84 & 3.41 & 3.06 & 3.06 & 2.46 & 2.45 & 1.66 & 1.64 \\
        &1000& 3.32 & 3.32 & 3.03 & 2.31 & 2.30 & 1.75 & 1.74 & 1.19 & 1.19 \\
    [0.5ex] \hline
    (1)&  10 & 3.85 & 4.41 & 3.38 & 3.06 & 3.53 & 2.47 & 2.88 & 1.67 & 2.03 \\
       &1000 & 3.31 & 3.48 & 3.0  & 2.30 & 2.42 & 1.74 & 1.84 & 1.19 & 1.26 \\
    [0.5ex] \hline
    (2)&  10 & 3.89 & 4.73 & 3.40 & 3.09 & 3.83 & 2.48 & 3.15 & 1.67 & 2.25 \\
       &1000 & 3.33 & 3.61 & 3.02 & 2.32 & 2.52 & 1.75 & 1.91 & 1.19 & 1.30 \\
    [0.5ex] \hline
    (3)&  10 & 3.86 & 6.44 & 3.38 & 3.05 & 5.28 & 2.44 & 4.44 & 1.64 & 3.34 \\
       &1000 & 3.32 & 4.19 & 3.02 & 2.31 & 2.93 & 1.74 & 2.24 & 1.20 & 1.56 \\
  [0.5ex] \hline
  \end{tabular}
  \end{center}
\end{table}
widths are given 
for distributions 
containing only muons above certain kinetic energy thresholds 
(including zero threshold, or no cut). In addition,
the corresponding values from the calculations without the effect of the 
magnetic field are given.
The widths in the North-down plane are much less affected than in the East-down 
plane which contains the dominant effect of the magnetic deflection.  
This effect increases with increasing values of $B_{x}$ or decreasing 
values of muon momentum.
In addition to the widths of the muon's sea level angular distributions, 
Table~\ref{angwidthsmag.tab} also shows the values
at their production vertex (birth, E-d.(b)).
Interestingly, these values are almost constant for the different
field conditions.  The effect of the magnetic field on the angular distributions
is therefore mainly the deflection of the muon over its longer path
and not the deflection of the parent pions or other charged particles 
preceding in the cascade 
due to their shorter path lengths and higher momenta as compared to
the muons.
In addition, the dominant factor in the final width (except for the highest 
magnetic field value) is the muon's angular distribution at its production 
(birth) point.  
%
%
\section{The effect of pressure and temperature variations}
\label{pressure.sec}
Air pressure and temperature changes cause small variations in the density 
profile of the atmosphere and thus affect the muon flux at sea level.
In order to obtain a quantitative estimate of these two effects,
the following cases were
studied: (i) the air density in each layer was increased by 3\% while all
other variables (geometry, cutoffs, etc.) were kept constant and (ii)
the height above sea level of each slab boundary was raised by 5\%
and the corresponding density decreased in the same proportion such 
that the total thickness of air (in g/cm$^2$) above the surface of the Earth
remained constant. 
The former modification simulates an air pressure increase ($p$) 
whereas the latter simulates an overall 5\% increase in absolute 
temperature ($T$) of the atmosphere.
Since muons are produced at atmospheric heights far above those
which determine the weather at the surface of the Earth, the mean $T$
which is involved in this calculation has little correlation with the
temperature at the Earth's surface.  
Thus the temperature effect calculated here 
is not expected to correlate with the Earth's
surface temperature but would require temperature measurements
of the Earth's atmosphere averaged from 0 to 20~km above the Earth's 
surface as obtained, for example, in weather balloon measurements.
The results presented in this section allow an investigation of, or 
corrections for, 
the muon flux due to small variations in the atmospheric pressure 
and height-averaged temperature.

The results for the total muon yield per primary gamma ray are summarized 
for two energies (10~GeV and 1~TeV) in Table~\ref{temp-pres.tab}. In all  
\begin{table}[htbp]
 \caption{\label{temp-pres.tab}
    Summary of the effect of pressure and temperature variations
    on the average muon multiplicity per primary gamma ray at detection level,
    $N_{\mu}/N_{\gamma}$. The column labeled ``ref. atmosphere'' 
    repeats the values listed in Table \protect\ref{no-of-muons.tab} for the 
    reference problem. 
    The ``5\% temperature increase'' corresponds to an 
    average increase in the absolute temperature profile from 0 to 80 km.
         }
  \vspace{2mm}
  \begin{center}
  \begin{tabular} {|c||c|c|c|}
    \hline
    $E_{\gamma}$
    & \multicolumn{3}{c|}{Muon multiplicity per primary gamma ray, 
                          $N_{\mu}/N_{\gamma}$}\\
    \hline
    (GeV) & ref. atmosphere & 3\% density increase & 5\% temperature increase \\
    [0.5ex] \hline \hline
    10 & $(5.29\pm 0.03)\times 10^{-4}$ & $(5.08\pm 0.02)\times 10^{-4}$
       & $(4.97\pm 0.02)\times 10^{-4}$  \\
  1000 & $0.221\pm 0.001$ & $0.214\pm 0.001$
       & $0.213\pm 0.001$ \\
  [0.5ex] \hline
  \end{tabular}
  \end{center}
\end{table}
cases the muon yields decreased by 3\% to 6\%, the decrease
being more pronounced at the lower energy. As for increased pressure, 
a higher density increases the interaction
probability of the pions and thus decreases the probability of decay, i.e.,
fewer muons are produced. As well, the added thickness of air which the 
muons must traverse is a barrier which some will fail to overcome.  
On the other hand, raising the boundaries of the 
layers and reducing their density 
causes two counterbalancing effects: (i) the pions propagate 
in a less dense medium and thus have a greater probability to decay 
rather than interact thus producing more muons, 
and (ii) the muons must propagate longer distances resulting in a greater 
probability that they will decay before reaching ground level.  
The calculated temperature dependence of the muon multiplicity ratio 
suggests that the enhanced 
muon decay probability dominates at these energies 
and is the more dominant at the lower energy.
%
%
\section{General properties of muon production in gamma ray showers}
\label{genprop.sec}
Information on the muons at sea level and on their ancestors (i.e.,
the parent, grandparent, etc., see 
Sec.~\ref{fluka-simdet.subsec}) allows a more detailed study of the properties
of these ancestors.
%
\subsection{The ancestors of the muons}
Table~\ref{decaymesons.tab} shows which parent contributes through its decay
to the detected muons for the different primary gamma ray energies. 
\begin{table}[htbp]
 \caption{\label{decaymesons.tab}
    Fractional contributions to the parents of the muons which 
    reach sea level.  
    Values are given for different energies of the primary cosmic gamma ray.
         }
 \vspace{2mm}
  \begin{center}
  \begin{tabular} {|c||c|c|c|c|c|}
  \hline
 $E_{\gamma}$ (GeV) & $\pi^+$ & $\pi^-$ & K$^+$ & K$^-$ & neutral kaons \\
    [0.5ex] \hline \hline
   1 & 0.106 & 0.894 & 0.0   & 0.0   & 0.0     \\
   3 & 0.495 & 0.485 & 0.020 & 0.0   & $1.7\times 10^{-4}$  \\
  10 & 0.492 & 0.489 & 0.011 & 0.007 & $9.8\times 10^{-4}$  \\
  30 & 0.482 & 0.482 & 0.019 & 0.014 & $3.1\times 10^{-3}$  \\
 100 & 0.478 & 0.477 & 0.022 & 0.018 & $4.4\times 10^{-3}$  \\
 300 & 0.477 & 0.476 & 0.023 & 0.019 & $4.7\times 10^{-3}$  \\
1000 & 0.475 & 0.476 & 0.024 & 0.019 & $5.2\times 10^{-3}$  \\
3000 & 0.476 & 0.475 & 0.025 & 0.020 & $5.1\times 10^{-3}$  \\
10000& 0.474 & 0.477 & 0.024 & 0.020 & $5.2\times 10^{-3}$  \\
  [0.5ex] \hline
  \end{tabular}
  \end{center}
\end{table}
At the lowest
energy (1~GeV) about 89\% of the muons are produced in decays of negative pions
and 11\% in decays of positive pions. This asymmetry can be explained by
the different interaction cross sections at low energy of pions of either 
charge and by the fact that these pions are mainly produced in secondary
interactions of neutrons (see below). Correspondingly,
the probability for positive pions to interact instead of decay is
larger than for negative pions. Above 3~GeV this asymmetry disappears and
decaying kaons begin to contribute to the sea level muon flux. The latter
contribution increases with energy and amounts to about 5\% at 10~TeV.

It is interesting to note that at all energies {\sc FLUKA} predicts the 
relative contribution of a positive kaon parent to be always larger than
that of a negative kaon by about 20\%. This effect is due to the properties of
the DPM describing inelastic hadronic interactions within {\sc FLUKA}.
In particular, it is a feature of the Reggeon contribution which describes
particle production by one quark-diquark string stretched between a 
valence quark of the fluctuating photon and a diquark
of a target nucleon. In this picture, kaon production involves the creation
of a $s\bar{s}$ quark-antiquark pair and, in case of negative kaons, also the
creation of an $u\bar{u}$ pair. On the other hand, positive kaons can be 
readily formed also by a $u$-quark of the fluctuating photon leading to the
observed asymmetry.

The fractional contribution to the muon's grandparent 
is given in Table~\ref{mother.tab}.
The calculations for
1~GeV show distinctly different features. About 97\% of the 
muons at sea level originate from mesons produced in interactions of 
neutrons in the close vicinity of the detector. At all other energies 
photoproduction
dominates the picture with a contribution decreasing with energy from 
99\% at 3~GeV to 60\% at the highest energy; the remaining fraction
is mainly from pions.
\begin{table}[htbp]
 \caption{\label{mother.tab}
    Fractional contributions to the grandparent 
     of the detected muon at sea level.  
         }
 \vspace{2mm}
  \begin{center}
  \begin{tabular} {|c||c|c|c|c|c|}
  \hline
 $E_{\gamma}$ (GeV) & $\gamma$ & p and $\bar{\rm p}$ & n and $\bar{\rm n}$ &
                      $\pi^+$ & $\pi^-$  \\
    [0.5ex] \hline \hline
   1 & $1.7\times 10^{-3}$ & 0.023               & 0.97
     & $7.8\times 10^{-4}$ & $4.9\times 10^{-3}$ \\
   3 & 0.99                & $3.9\times 10^{-4}$ & $5.0\times 10^{-3}$
     & $2.3\times 10^{-3}$ & $2.1\times 10^{-3}$ \\
  10 & 0.99                & $8.2\times 10^{-4}$ & $1.9\times 10^{-3}$
     & $2.6\times 10^{-3}$ & $2.7\times 10^{-3}$ \\
  30 & 0.97                & $4.1\times 10^{-3}$ & $5.3\times 10^{-3}$
     & $9.3\times 10^{-3}$ & $9.2\times 10^{-3}$ \\
 100 & 0.91                & 0.014               & 0.015             
     & 0.026               & 0.027               \\
 300 & 0.83                & 0.025               & 0.026             
     & 0.056               & 0.056               \\
1000 & 0.73                & 0.037               & 0.041             
     & 0.091               & 0.090               \\
3000 & 0.65                & 0.049               & 0.049             
     & 0.12                & 0.12                \\
10000& 0.60                & 0.054               & 0.057             
     & 0.13                & 0.14                \\
  [0.5ex] \hline
  \end{tabular}
  \end{center}
\end{table}
%
\subsection{Distributions of the number of generations}
The distributions of the generation number of the grandparent
of the sea level muon is shown in Fig.~\ref{dNdg}. The lower the energy
of the primary gamma ray the smaller is the atmospheric shower resulting
in a relatively narrow distribution. This distribution is peaked at
generation one for energies below a few hundred GeV where the parent
is mainly produced in a photoproduction process of the
primary photon (see also Table~\ref{mother.tab}). At higher energies 
secondary hadron interactions contributes significantly
to the production of the parent. These secondary hadrons are mainly of
third or fourth generation and cause a shift of the peak of the total 
distribution at TeV energies. The large tail at these energies which extends
up to more than 100 generations reflects photoproduction processes of 
secondary photons in the large electromagnetic shower.
%
\subsection{The muon production heights}
The distributions of heights above sea level at which the detected muons were 
produced are shown for four primary energies in Fig.~\ref{dNdh}. 
As expected, at high energy there is 
a decrease in height with increasing energy of the primary gamma ray
because of the logarithmic increase of the shower length.
However, as the energy of the primary gamma ray decreases below 10~GeV,
the energy of the muon parent is close 
to the minimum energy required for the muon to penetrate the blanket of
air between its production and sea level. Hence, the most probable height
of muon production begins to decrease with energies decreasing below 10~GeV.  
The blip-up near zero height, most pronounced in the 3 and 10~GeV
distributions, represents the excess contribution from very low energy muons.
%
\subsection{Radial distributions}
The radial distributions of the muons at their production vertices and at sea
level are shown in Fig.~\ref{dNdA} for five different gamma ray energies. 
The radial distance, $R$, is defined with
respect to the shower axis ($z$ axis). The quantity $dN/dA$ denotes the number 
of muons per unit area and per primary gamma ray. The
radial distributions of the production vertices are labeled ``production.''  

All distributions extend to more than 10~km. Whereas the sea level distributions
have a relatively flat shape below $R=2$~km the production vertex
distributions are increasing toward the shower axis and exhibit a change in 
slope or discontinuity at about 500~m which is most pronounced at low primary 
energy. This discontinuity indicates that
two components with different shapes contribute. In order to further investigate
this feature the muon production vertices were scored separately for
those muons which originated from mesons produced in photoproduction processes
and those from mesons produced by interactions of nucleons or other hadrons.
The contributions from these components are shown for a primary energy of
10~GeV in Fig.~\ref{dNdA10GeV}. As can be clearly seen in the production vertex
distributions (Fig.~\ref{dNdA10GeV}a) the photoproduction component dominates
at small radii and the component due to interacting nucleons constitutes
the tail at large $R$. 
The difference is smaller at sea level (Fig.~\ref{dNdA10GeV}b) due
to multiple scattering of the muons in air.
%
\subsection{Energy distribution of the muons}
The kinetic energy distributions of muons at sea level and at the production 
vertex of these 
muons are shown in Fig.~\ref{dNdEkin}. Note, that the distributions are
multiplied by the muon energy $E_\mu$ in order to enhance possible spectral
structures at the higher energies.  
Whereas the sea level spectra are a smooth function with energy,
the distributions of the muon energies at their birth exhibit a
two-component structure.  
This two-component structure is again due to the 
superposition of muons from mesons generated in photoproduction processes
and those from interacting hadrons. The two contributions are
plotted separately in Fig.~\ref{dNdEkin10GeV} for 10~GeV primary photon energy.
It is clear that muons from photoproduction processes 
have a higher average energy.
%
%
\section{Conclusions}
\label{conclude.sec}
Air showers caused by cosmic gamma rays with energies below 10~TeV
were simulated using the Monte Carlo code {\sc FLUKA}. The primary 
gamma rays were assumed to enter the atmosphere vertically.
The reliability of {\sc FLUKA} predictions in this energy region 
was explored by comparing its predictions for proton- and 
nuclei-induced air showers with experimental data; 
good agreement was found.

Many general properties of muon production in atmospheric gamma ray 
showers were studied.  
As expected, most of the muons which reach sea level
are decay products of charged pions; the contribution from kaons increases
as the primary energy increases.
Below 100~GeV primary energy, the decaying mesons are produced mainly in
photoproduction processes.  At higher
energies, hadronic interactions of nucleons and pions contribute
significantly.
Variations in atmospheric 
pressure and temperature were investigated:  
A 3.0\% increase in atmospheric pressure decreases the muon yield at sea 
level by 4.0\% for 10~GeV primaries (3.2\% at 1~TeV).  Similarly,
a 5.0\% increase in the height-averaged temperature decreases 
the muon yield by 6.0\% (3.6\% at 1~TeV).

The main goal of the study was to determine the angular resolution for 
primary cosmic gamma rays from measurements of secondary muon angles at 
ground level. The angular distribution 
of muons at sea level narrows as the muon energy increases since 
low energy muons constitute the tails of the angular distributions.  
Thus, the angular resolution can be improved by eliminating the lower 
energy muons.  
The half-widths at half-maximum-height of the muon's projected angles
are $\le 1.1^\circ$ above 100~GeV but rise to a value of $3.2^\circ$ 
as the primary gamma ray energy is lowered to 10~GeV in the absence 
of the Earth's magnetic field deflection.
Primary energies below 10~GeV essentially do not contribute
muons at ground level.
The effects of magnetic bending
were considered separately as they depend on several factors; 
for example, they are more important in the East-down 
plane, for low primary energies, low muon energies, 
and larger North-components of the Earth's magnetic field.
 
The width of the angular distribution can be narrowed if 
the correlation of $\Theta_{xz}$ (angle projected onto the North-down plane)
with $x$ is taken into consideration (assuming the location of the
core of the shower is measured).  This narrowing can 
be accomplished either by eliminating large $x$ and $y$ values from the 
data or by compensating the measured angles knowing their expected mean 
values versus $x$ and $y$ as calculated from {\sc FLUKA}.

Finally, the average muon angular correlation was calculated for a 
spectrum of primary gamma rays characterized by a spectral index.  
For a soft spectrum, e.g. a differential spectral index $\alpha = 3.0$,
the average projected angular width (HWHM) is $2.6^\circ$;
for a mean value, $\alpha = 2.41$, the width is $1.6^\circ$.
If the spectrum is harder, e.g. $\alpha \leq 2.0$, the angular width
improves to a constant value of $1.0^\circ$.  Corresponding 
values for a space angle resolution ($\delta\Theta_{R}$(HWHM)) 
are $\sqrt{2}$ larger.  
As mentioned above, if data with larger $x$ and $y$ values
and/or the lower muon energies are removed, this angular correlation can
be improved. 
This precise calculation of the angular correlations provides 
experiments which study the angular location of primary gamma ray sources
by measuring muon angles with information on the 
angular resolution which is difficult to obtain experimentally.  
%
%
\section*{Acknowledgments}
Part of this work was supported by the Department of Energy under contract
DE-AC03-76SF00515.
Project GRAND is funded through grants from the University of
Notre Dame and private donations.
%
%

%
%
\newpage
\section*{Figure Captions}
\begin{figure}[tb]
 \hspace*{0mm}\epsfig{file=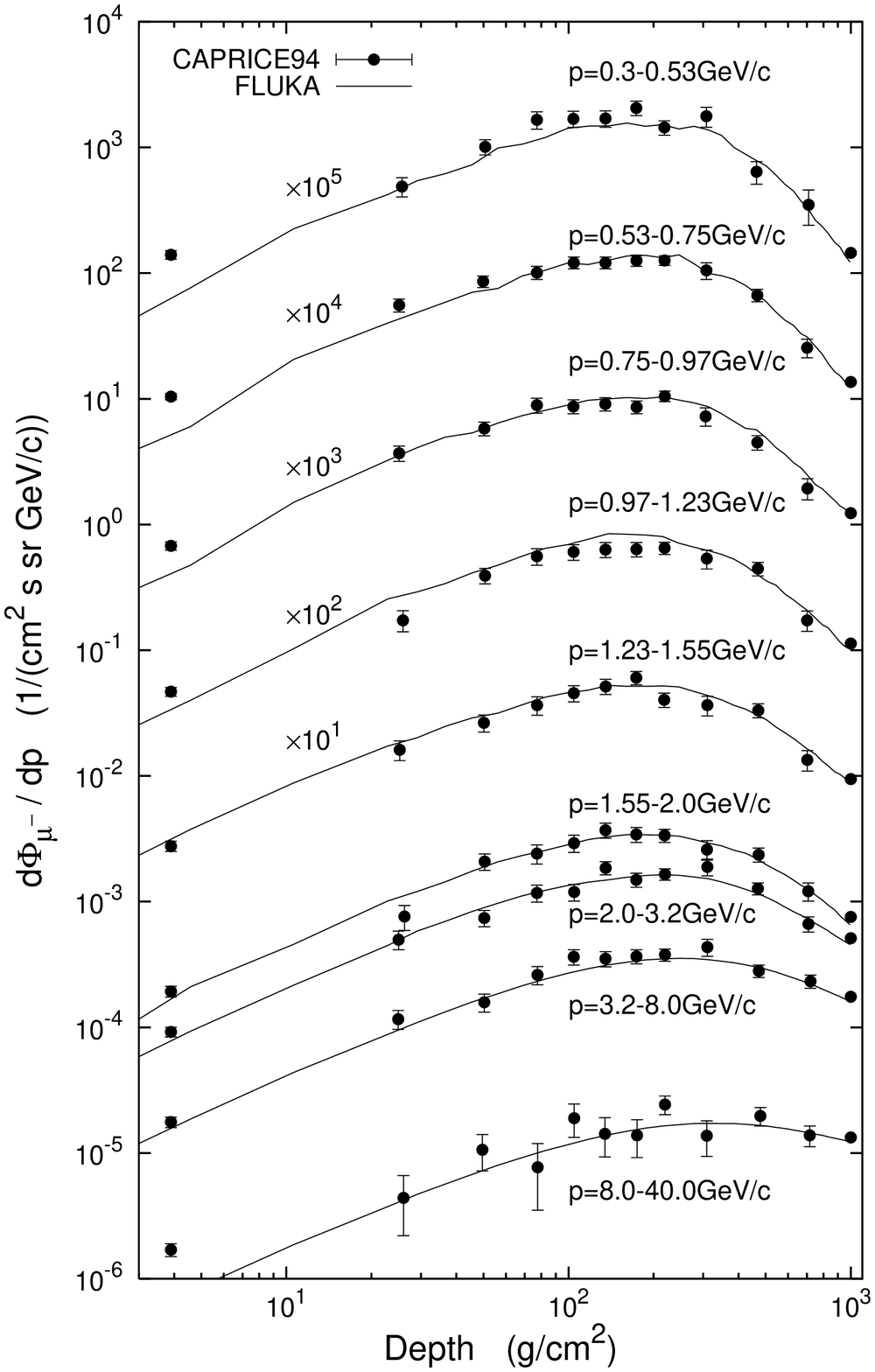, width=12cm}
 \caption{
         }
 \label{mu-capdep}
\end{figure}

\begin{figure}[htb]
 \hspace*{10mm}\epsfig{file=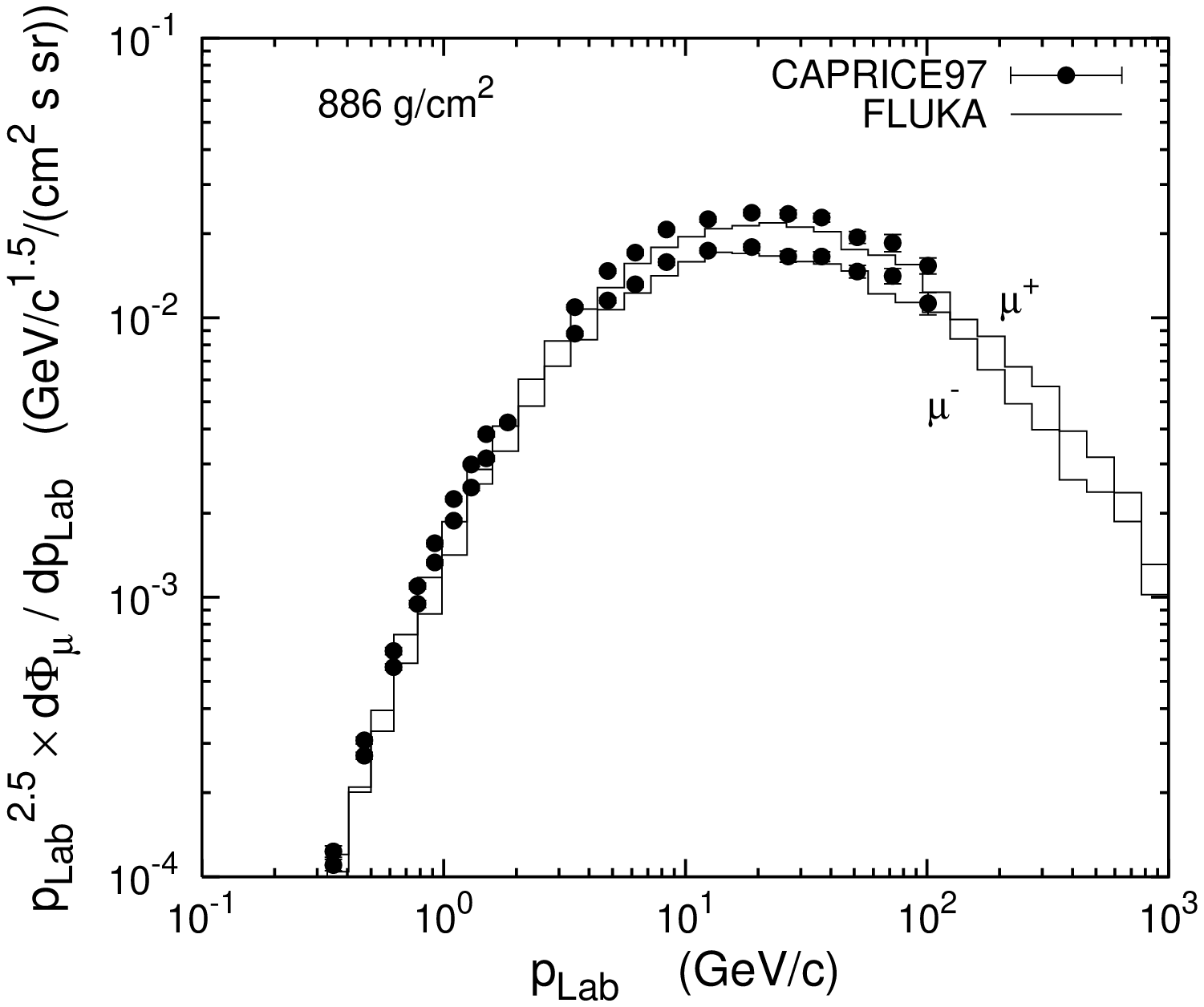, width=110mm}\\
 \hspace*{10mm}\mbox{(a)}\\
 \hspace*{10mm}\epsfig{file=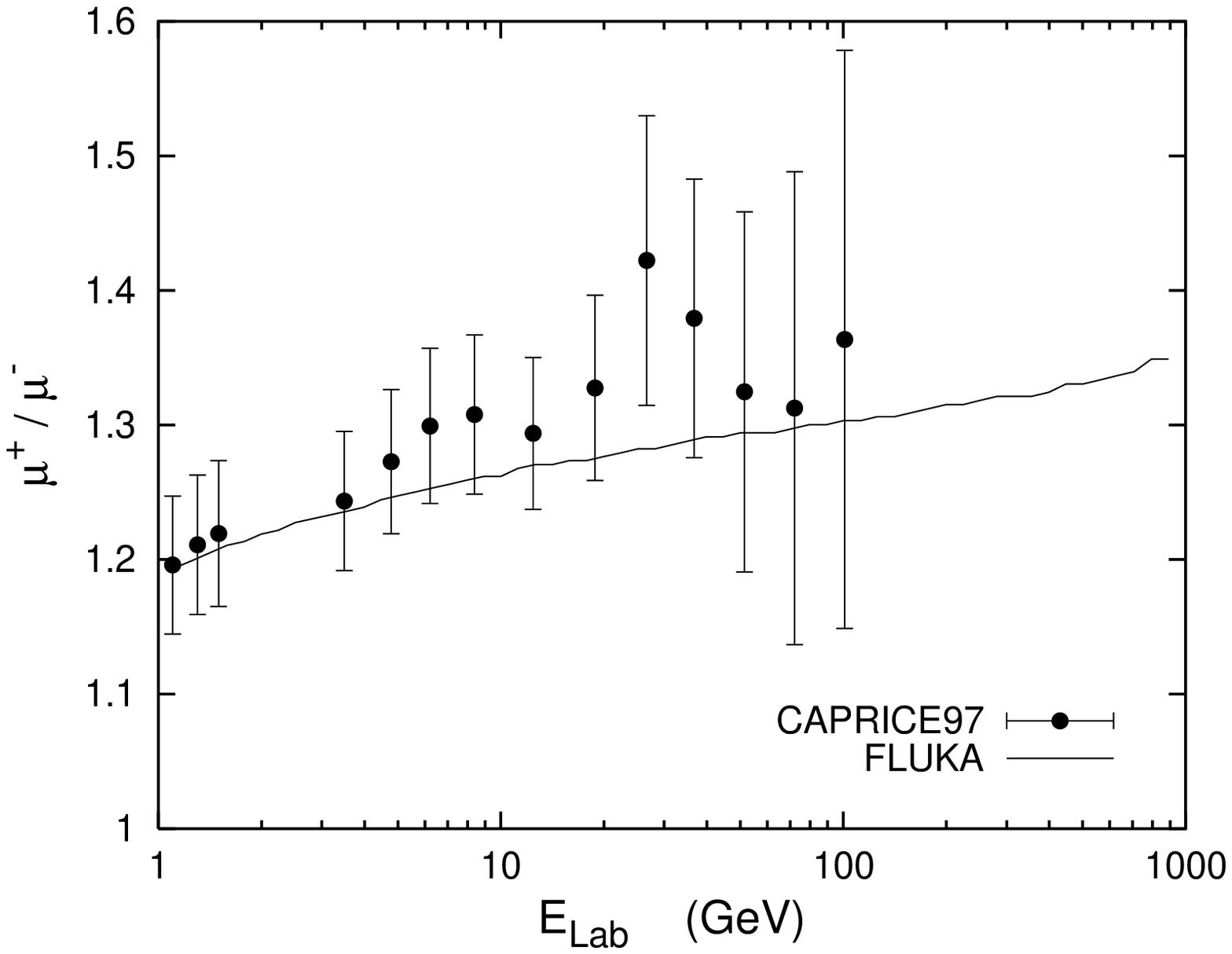, width=110mm}\\
 \hspace*{10mm}\mbox{(b)}
 \caption{
         }
 \label{mu-capspc25&ratio}
\end{figure}

\begin{figure}[htb]
 \hspace*{0mm}\epsfig{file=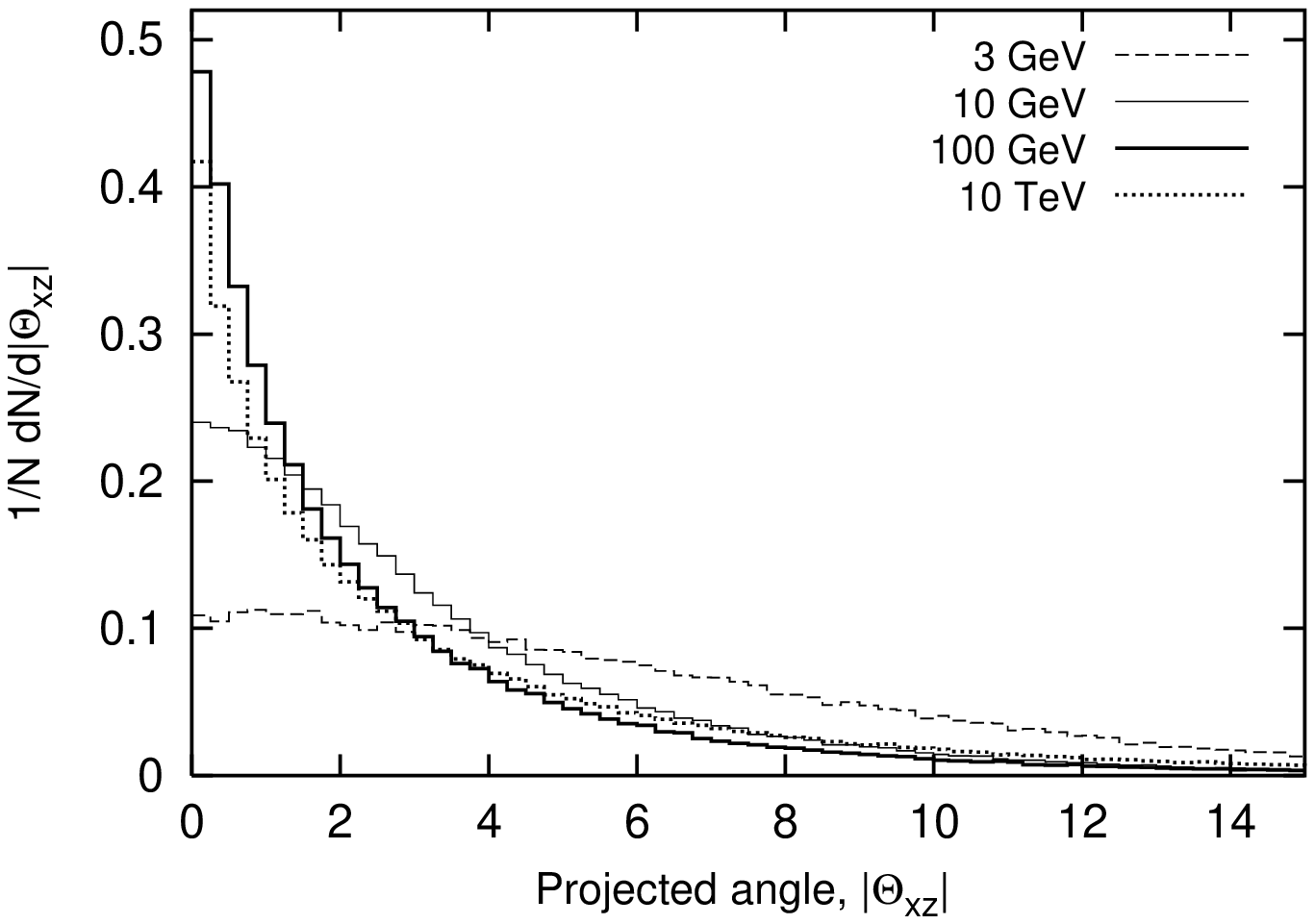,width=13cm}
 \caption{
         }
 \label{dNdTh}
\end{figure}

\begin{figure}[htb]
 \hspace*{10mm}\epsfig{file=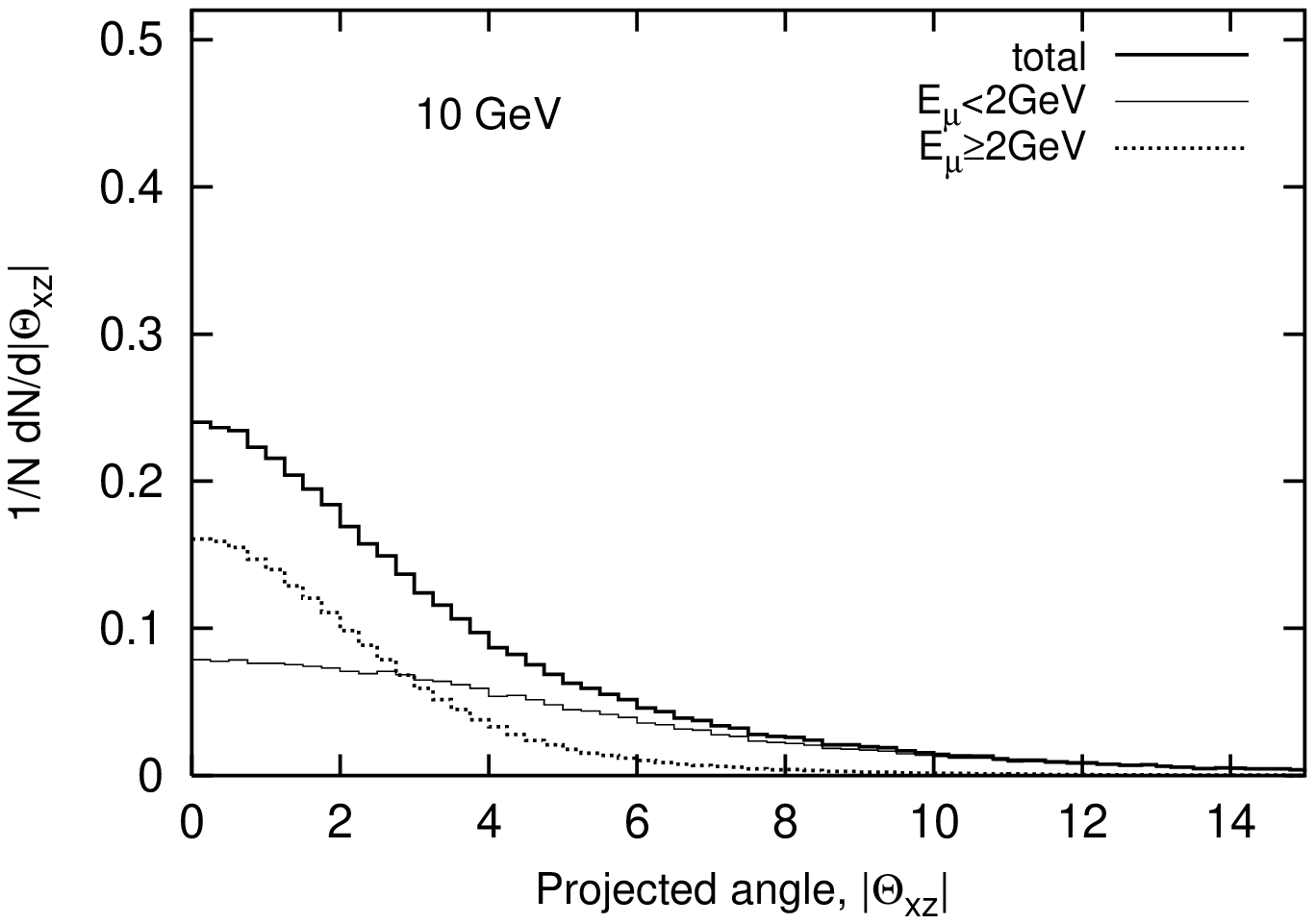,width=110mm} \\
 \hspace*{10mm}\mbox{(a)}\\
 \hspace*{10mm}\epsfig{file=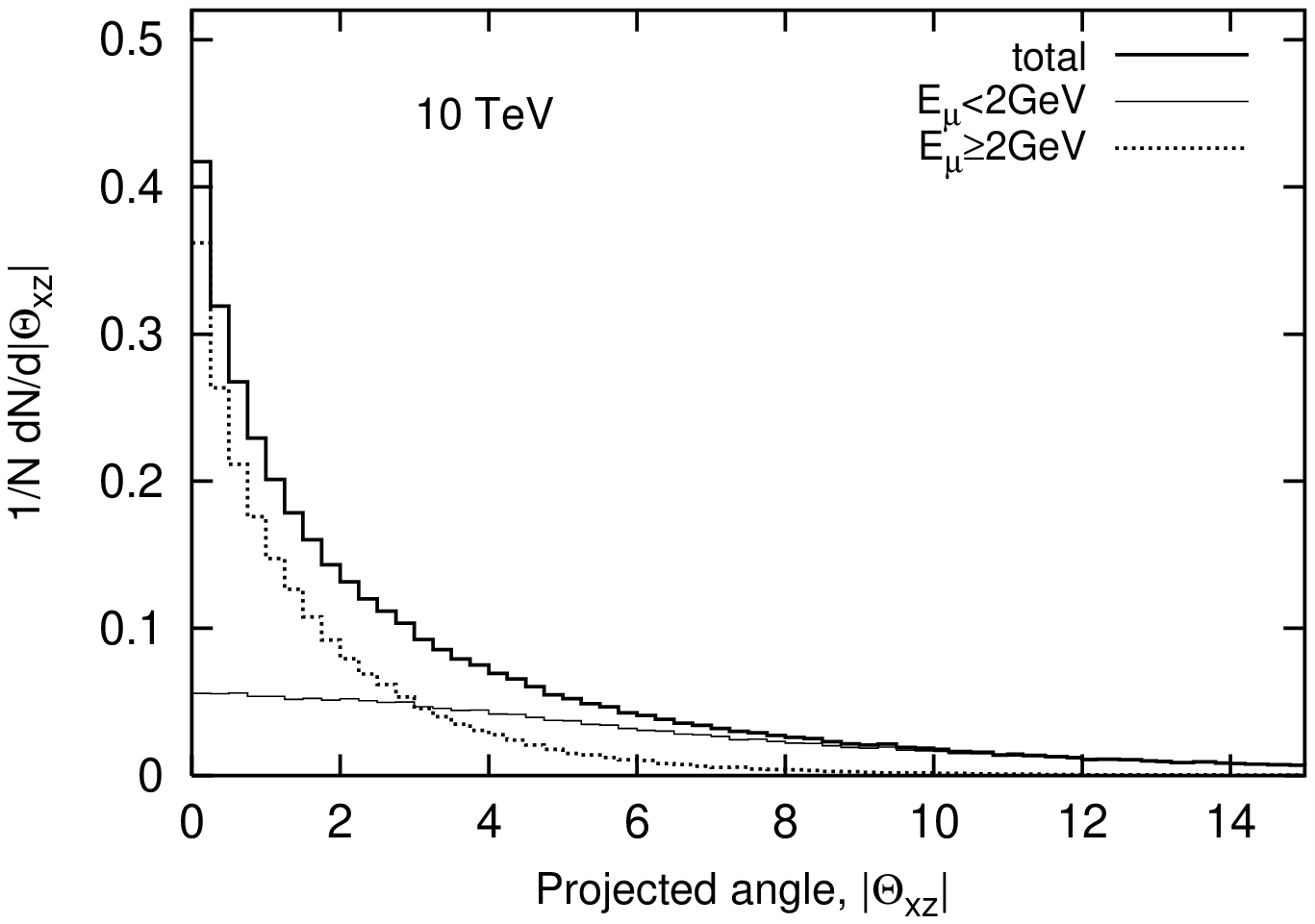,width=110mm} \\
 \hspace*{10mm}\mbox{(b)}
 \caption{
         }
 \label{dNdTh-10GTeV}
\end{figure}

\begin{figure}[htb]
 \hspace*{10mm}\epsfig{file=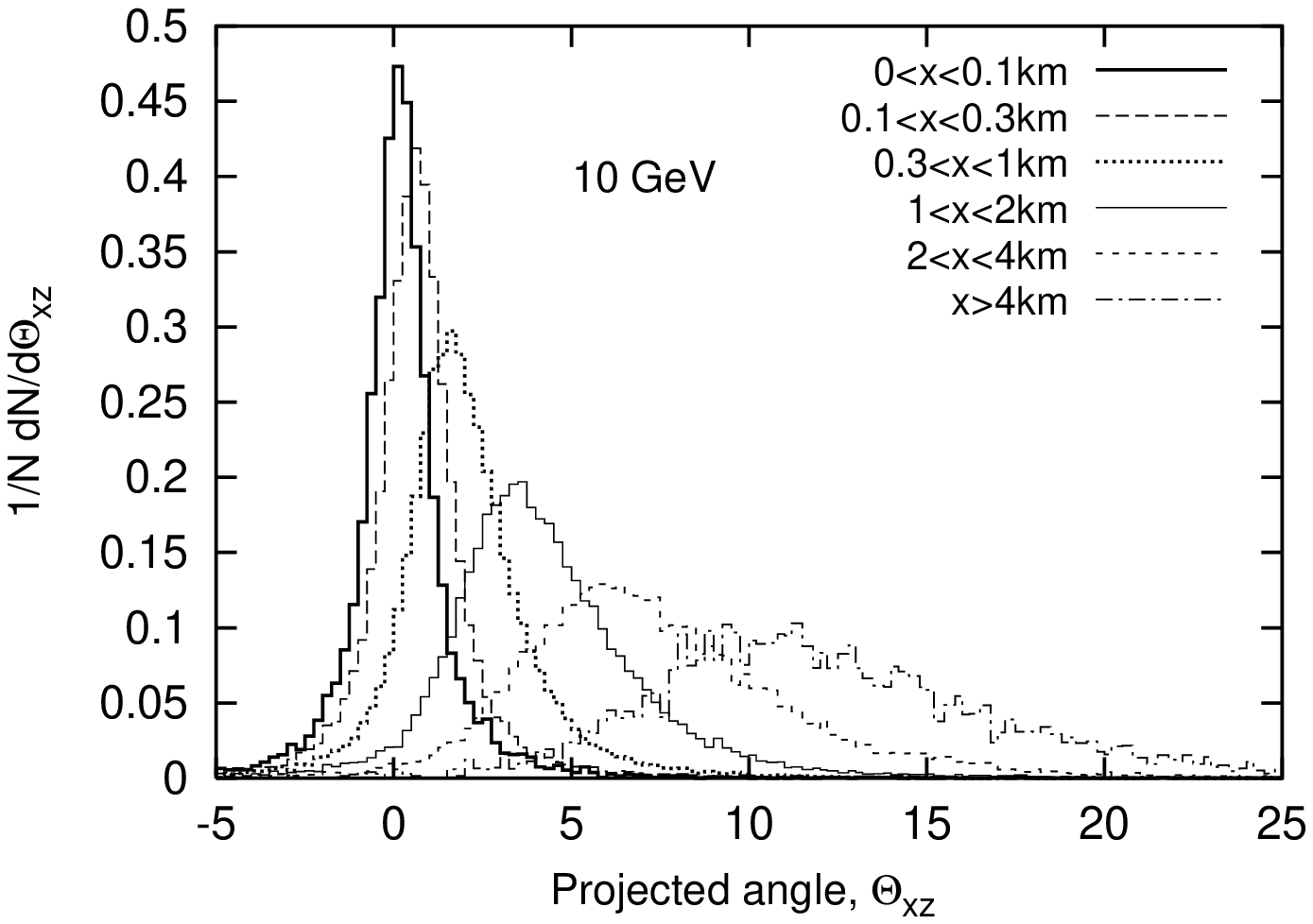,width=110mm} \\
 \hspace*{10mm}\mbox{(a)}\\
 \hspace*{10mm}\epsfig{file=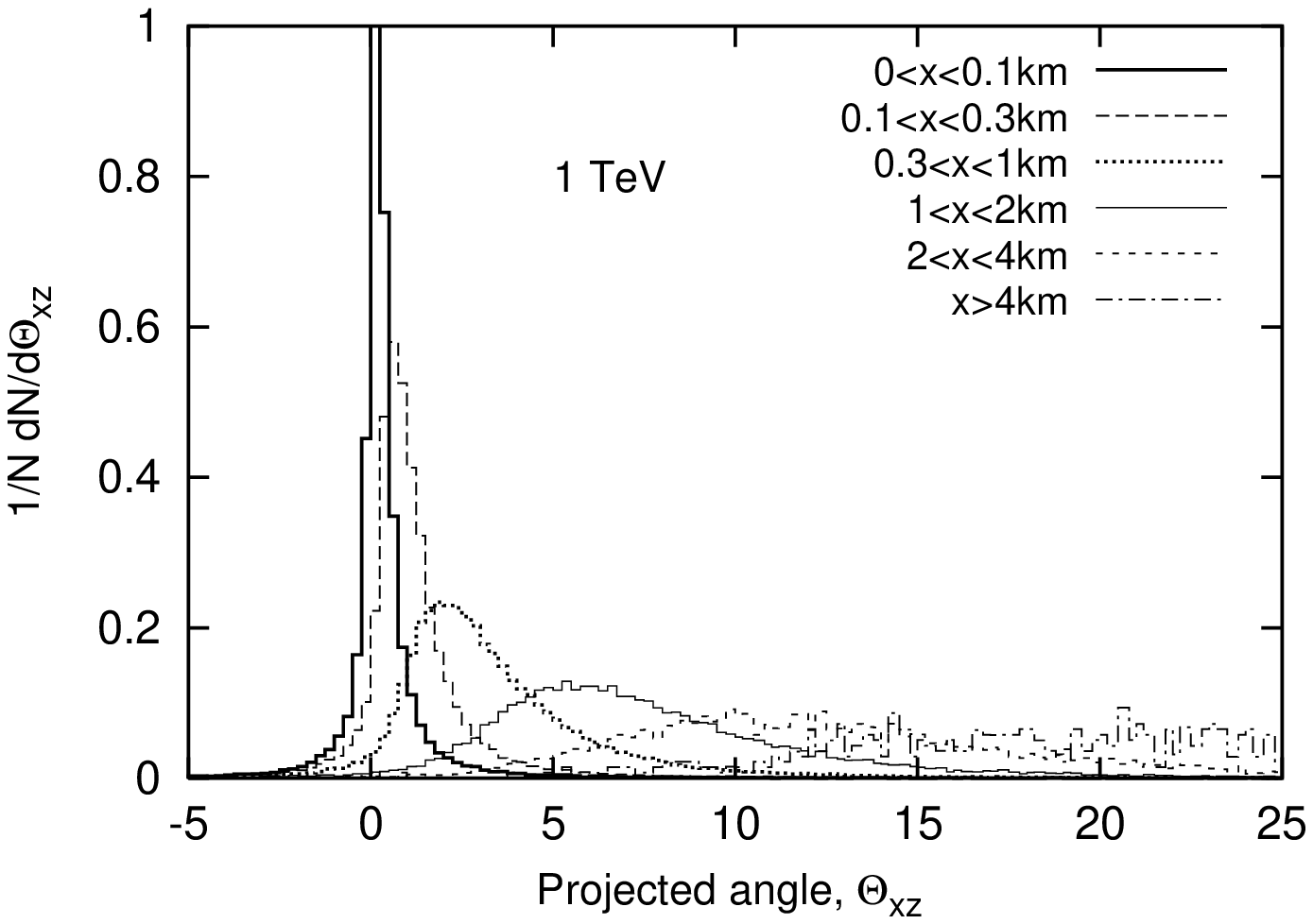,width=110mm}  \\
 \hspace*{10mm}\mbox{(b)}
 \caption{
         }
 \label{dNdThx-10GeV1TeVx}
\end{figure}

\begin{figure}[htb]
 \hspace*{-10mm}\epsfig{file=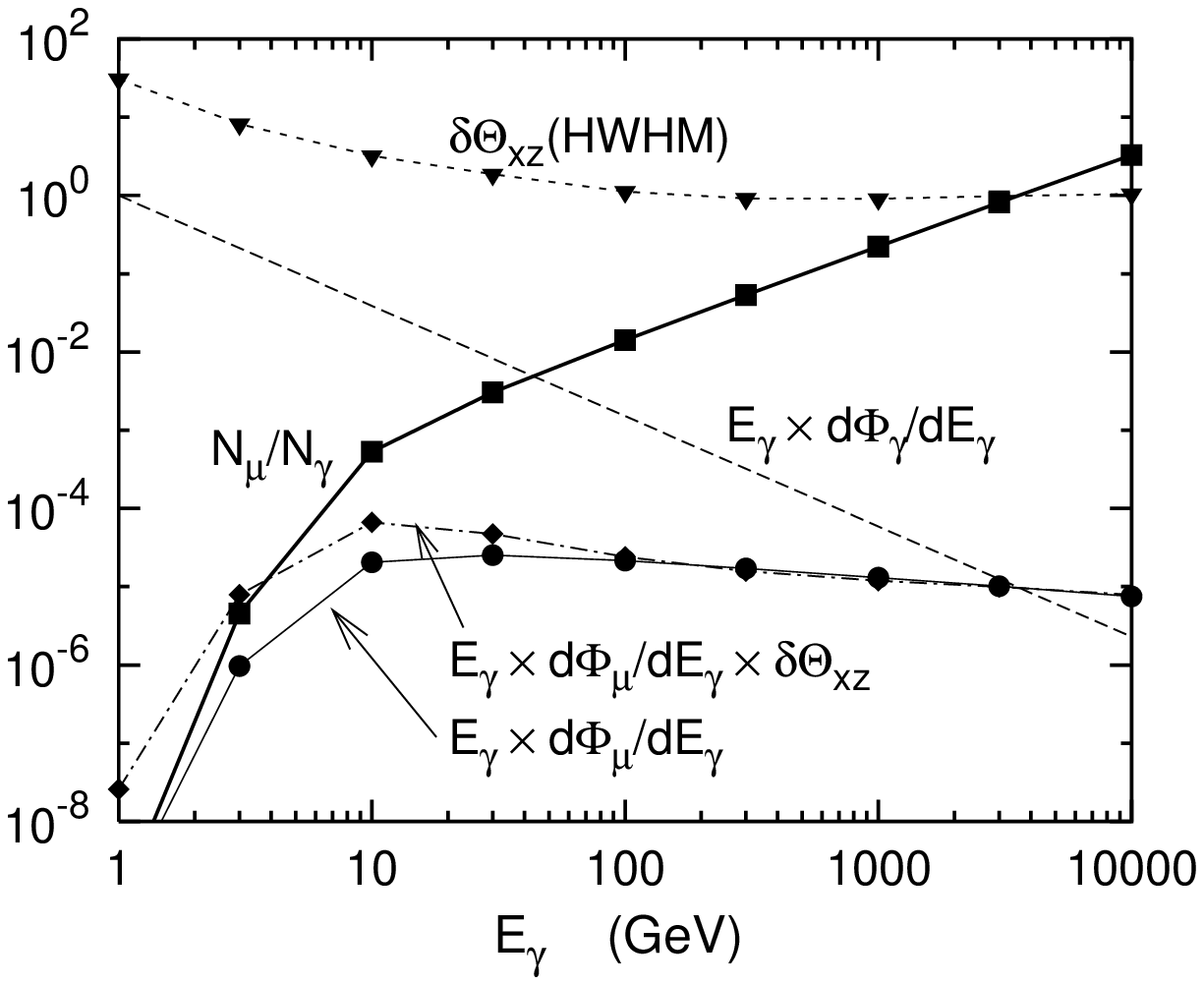,width=133mm} \\
 \hspace*{13mm}\mbox{(a)}\\
 \hspace*{13mm}\epsfig{file=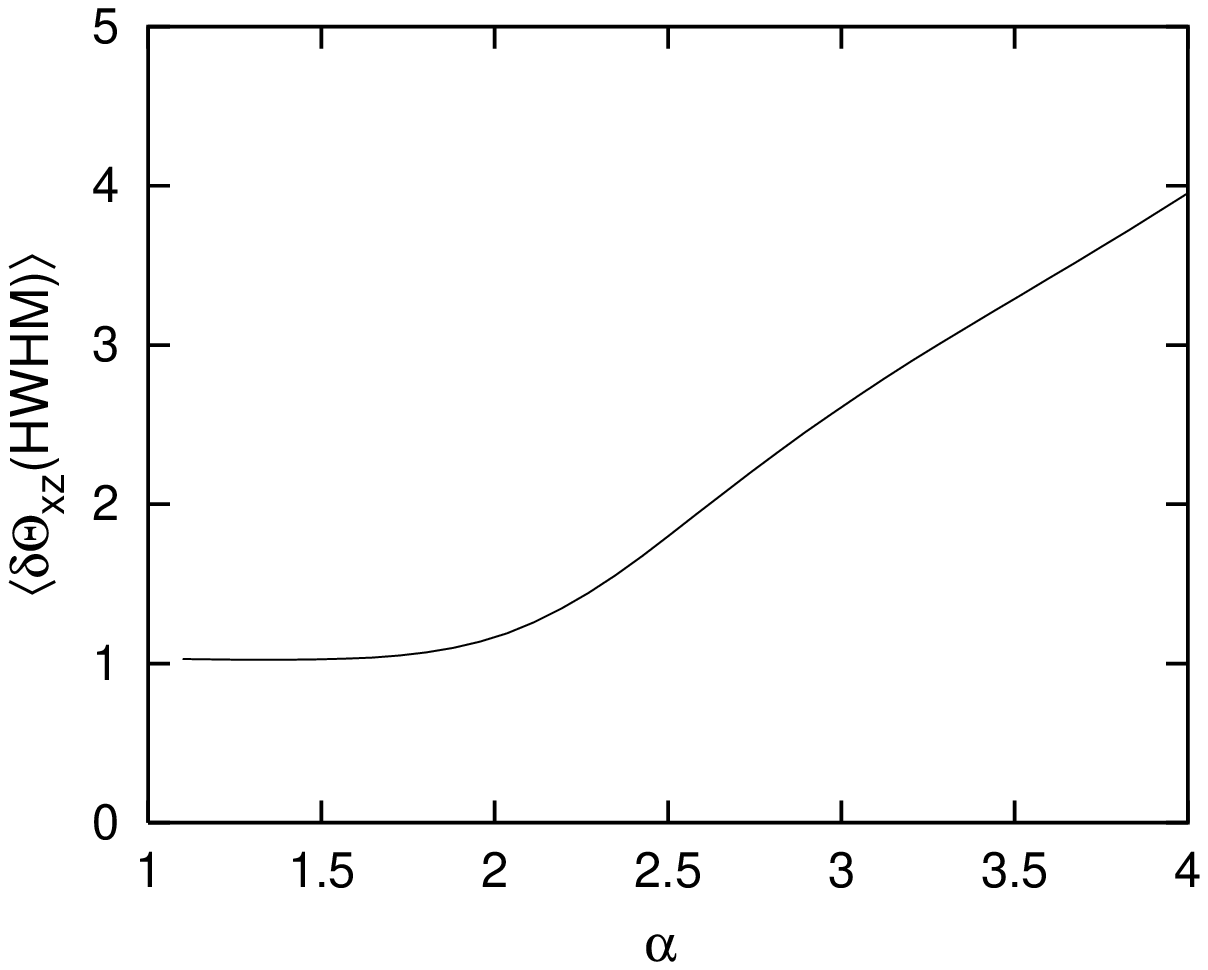,width=110mm} \\
 \hspace*{13mm}\mbox{(b)}
 \caption{
         }
 \label{flxgamma}
\end{figure}

\begin{figure}[htb]
 \hspace*{10mm}\epsfig{file=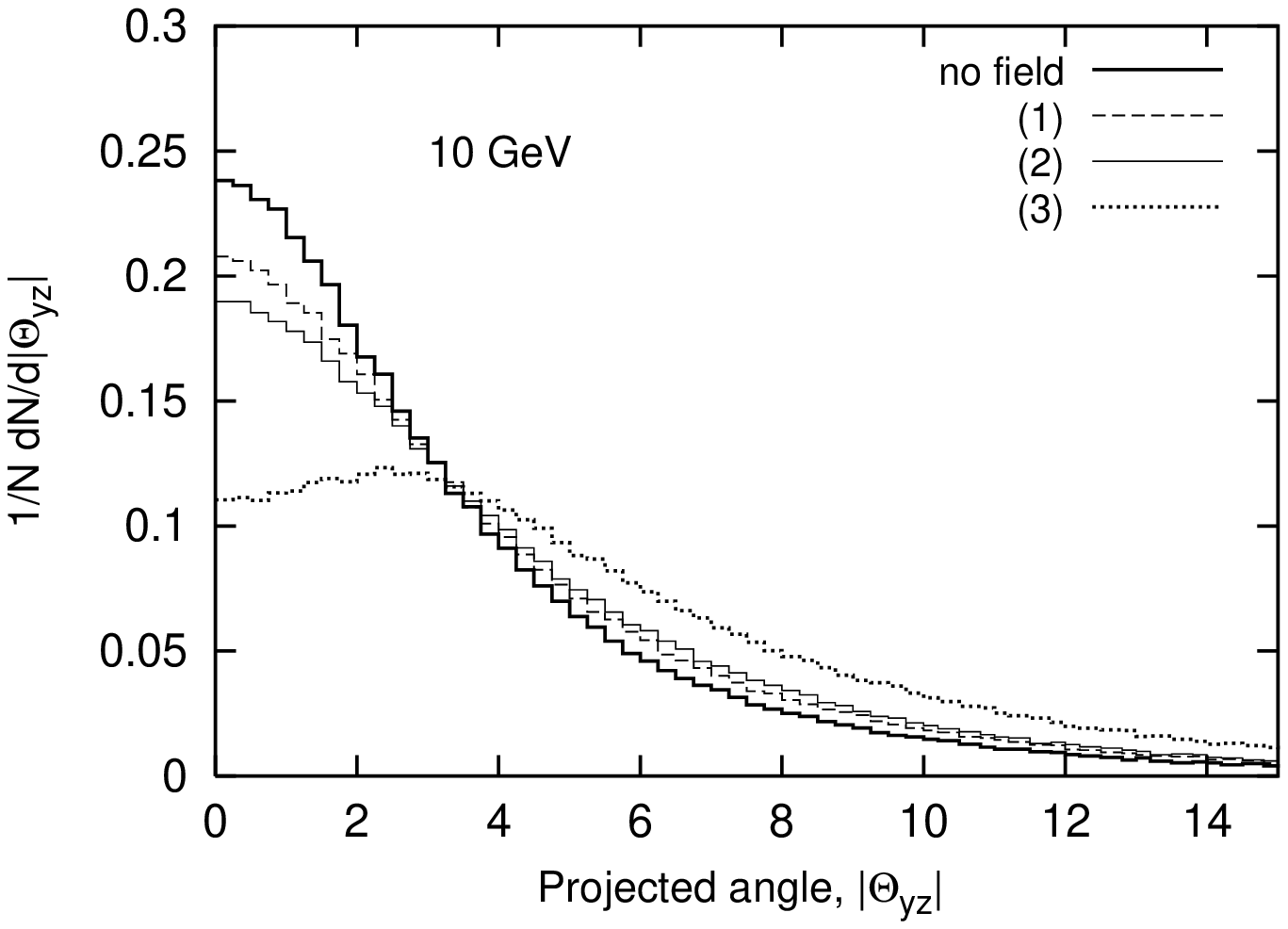,width=110mm} \\
 \hspace*{10mm}\mbox{(a)}\\
 \hspace*{10mm}\epsfig{file=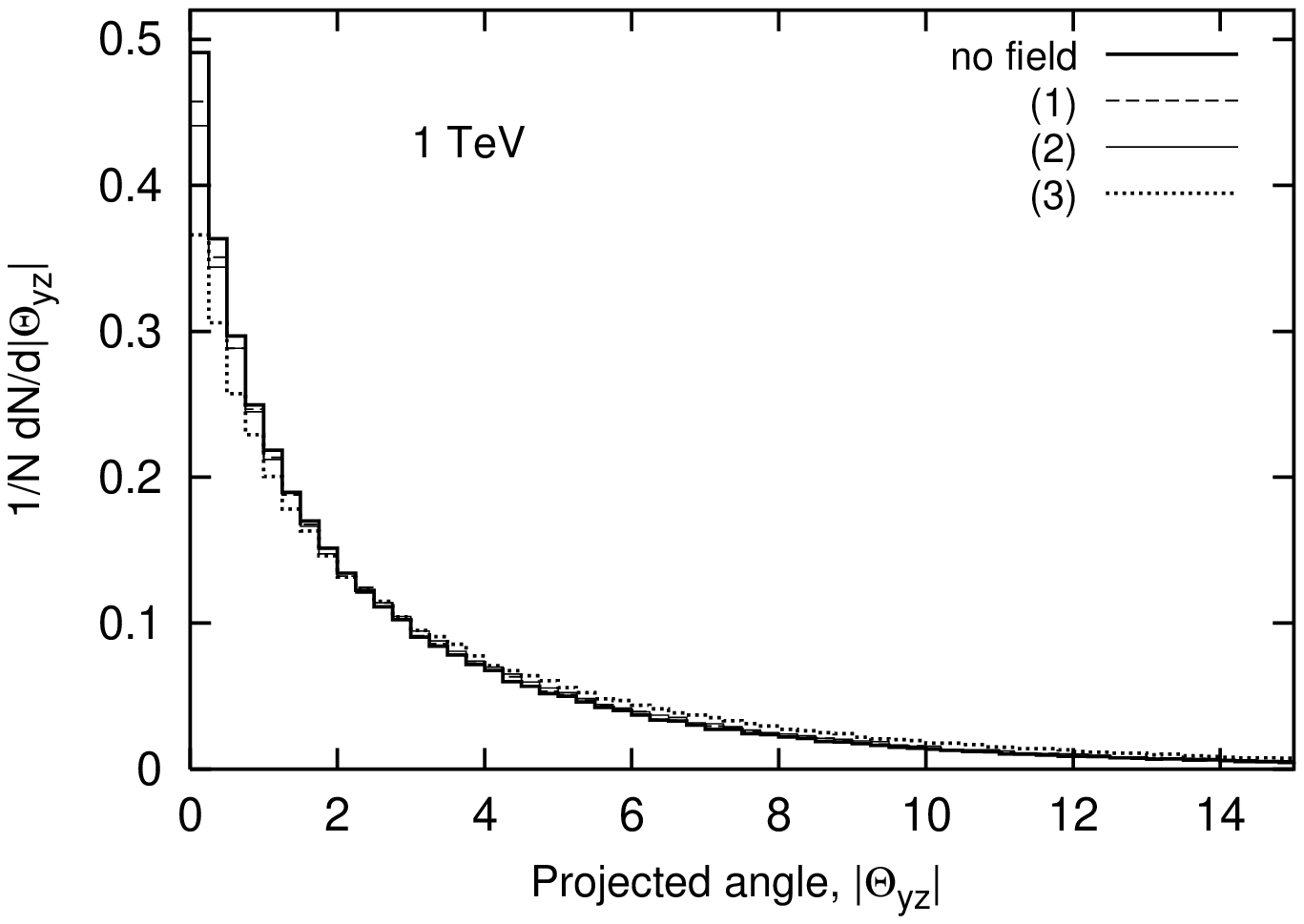,width=110mm} \\
 \hspace*{10mm}\mbox{(b)}
 \caption{
         }
 \label{dNdTh-10GeV1TeVmag}
\end{figure}

\begin{figure}[htb]
 \hspace*{-10mm}\epsfig{file=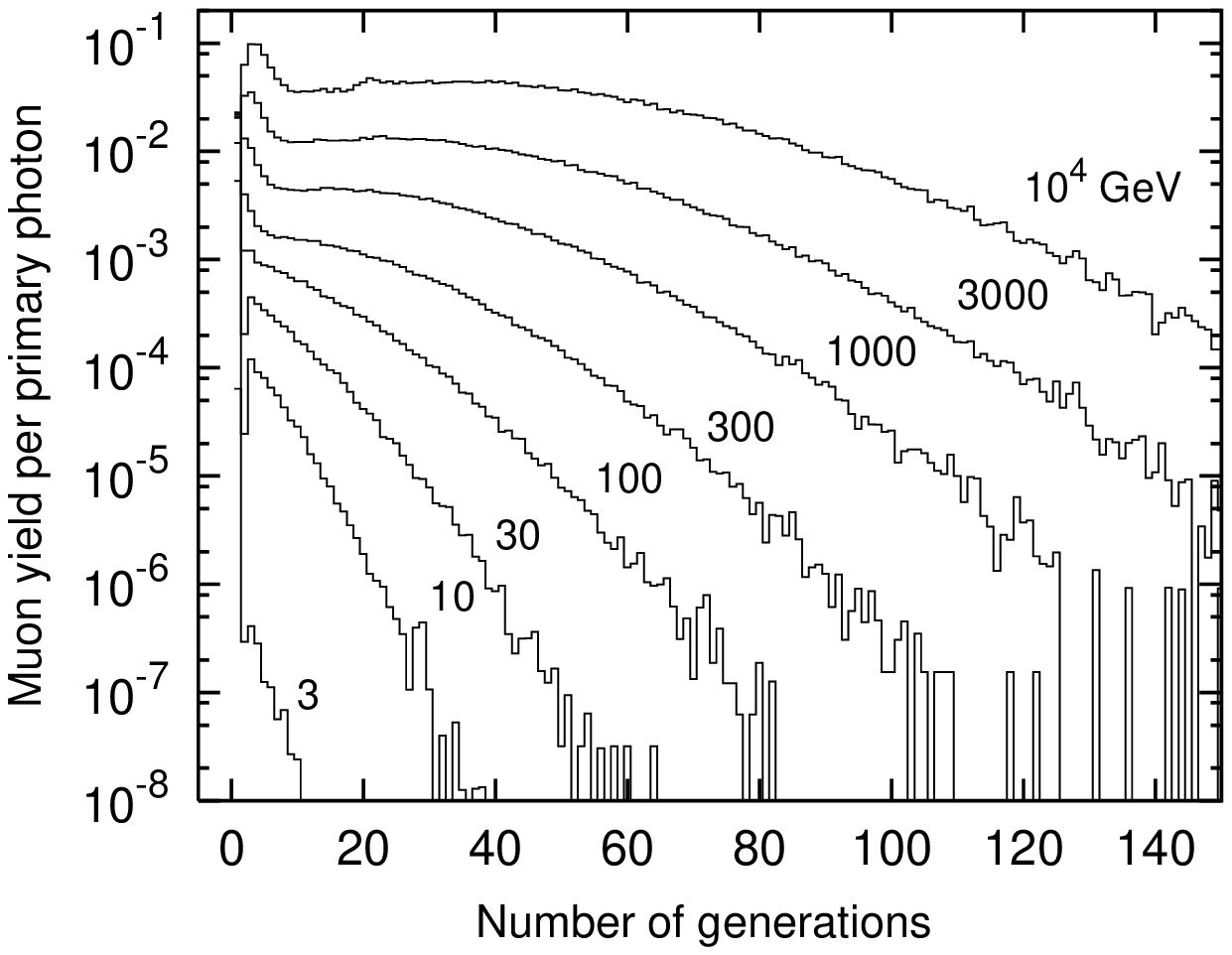,width=13cm}
 \caption{
         }
 \label{dNdg}
\end{figure}

\begin{figure}[htb]
 \vspace*{5cm}
 \hspace*{-10mm}\epsfig{file=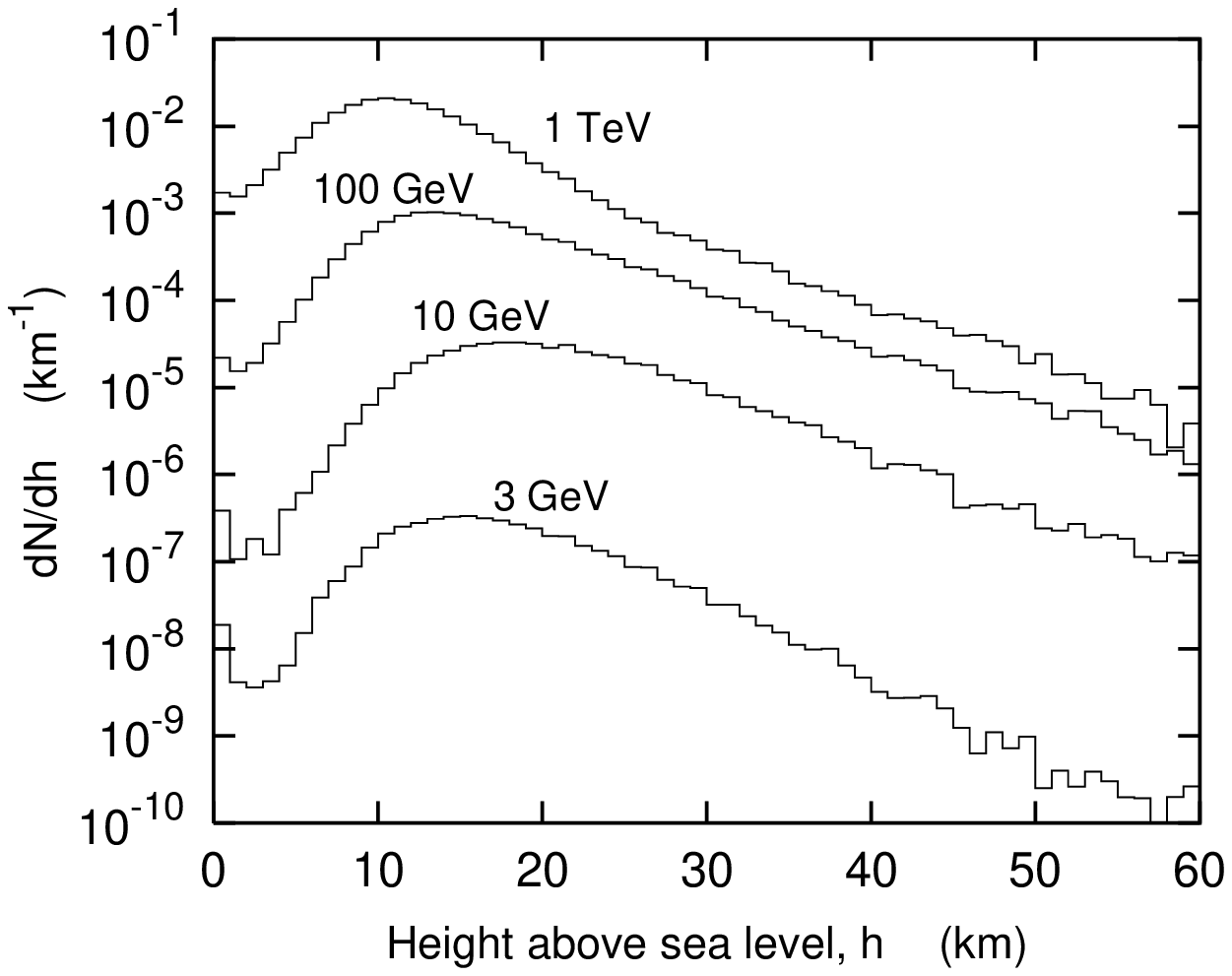,width=13cm} 
 \caption{
         }
 \label{dNdh}
\end{figure}

\begin{figure}[htb]
 \vspace*{5cm}
 \hspace*{-10mm}\epsfig{file=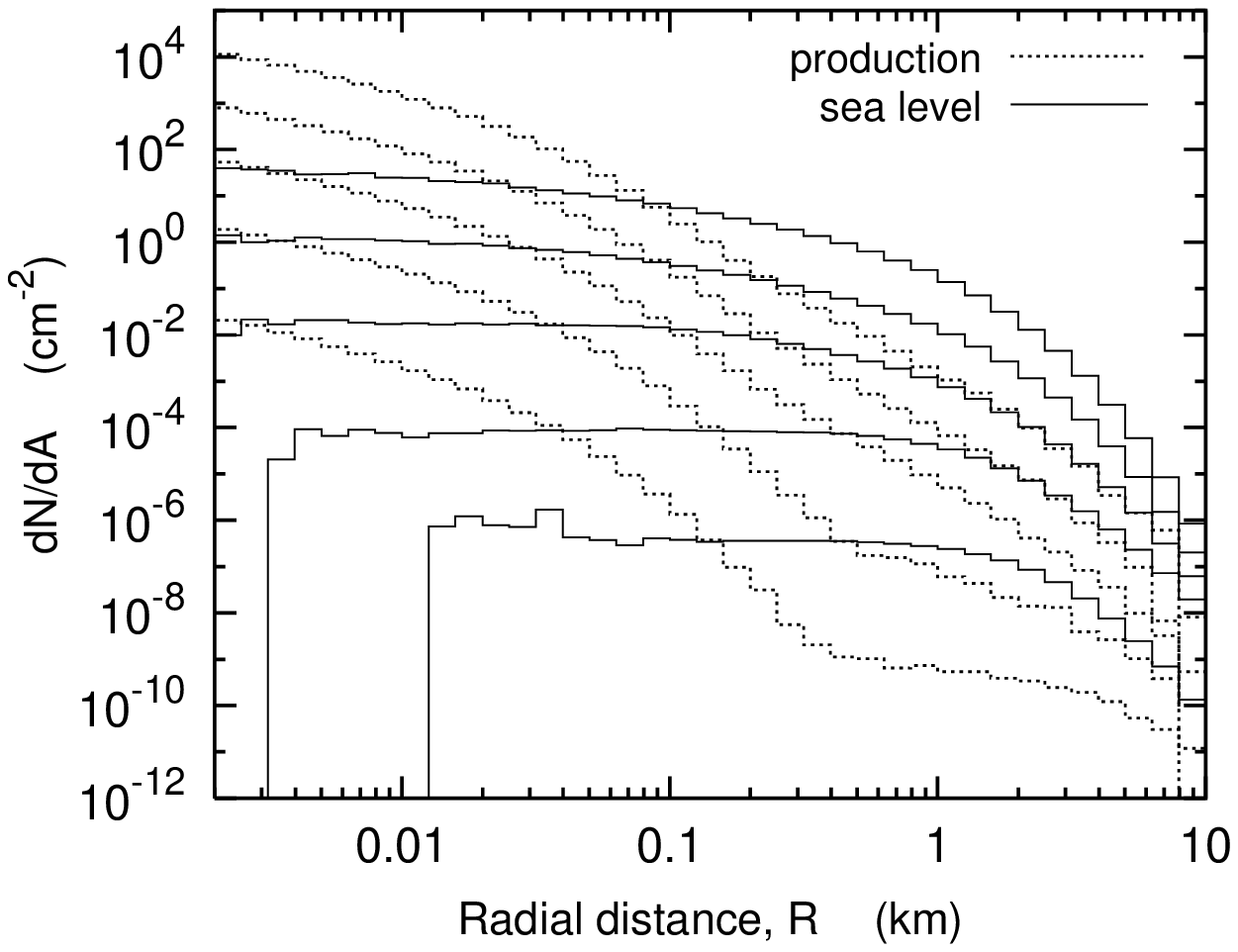,width=13cm}
 \caption{
         }
 \label{dNdA}
\end{figure}

\begin{figure}[htb]
 \vspace*{3cm}
 \hspace*{10mm}\epsfig{file=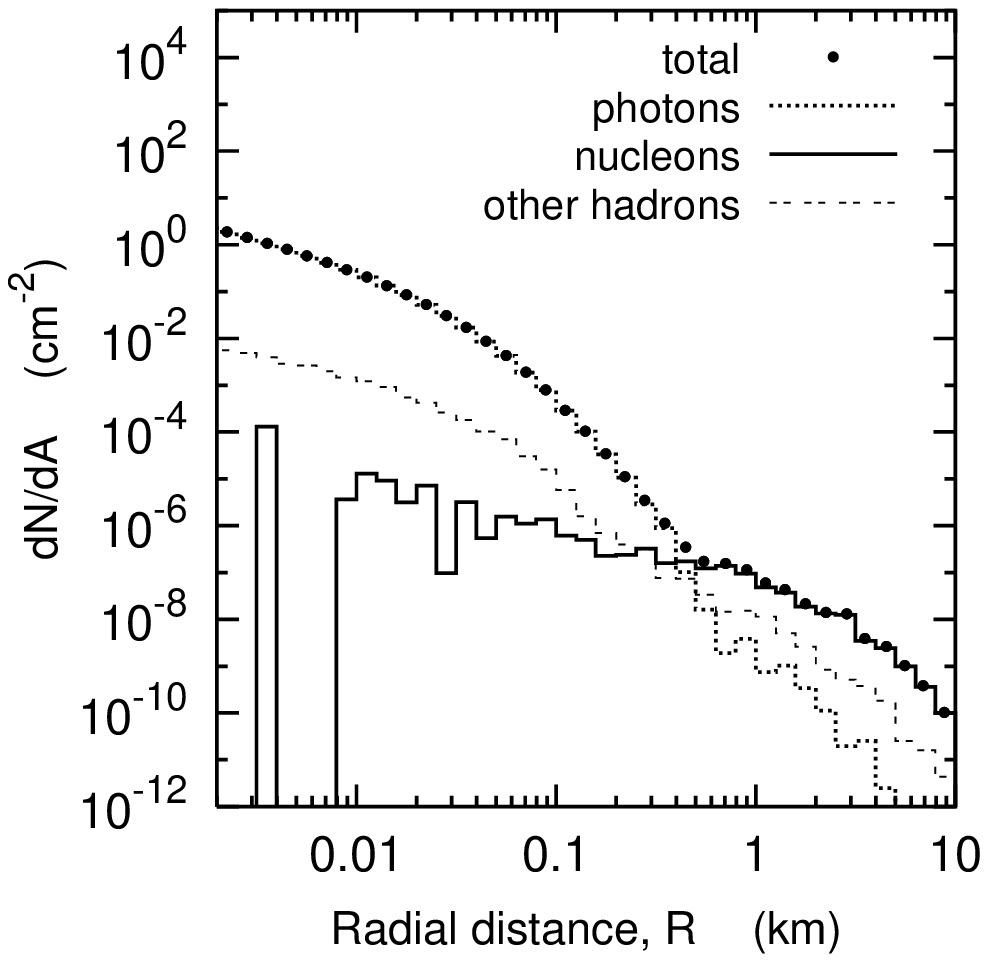, width=100mm}  \\
 \hspace*{20mm}\mbox{(a)}\\
 \hspace*{10mm}\epsfig{file=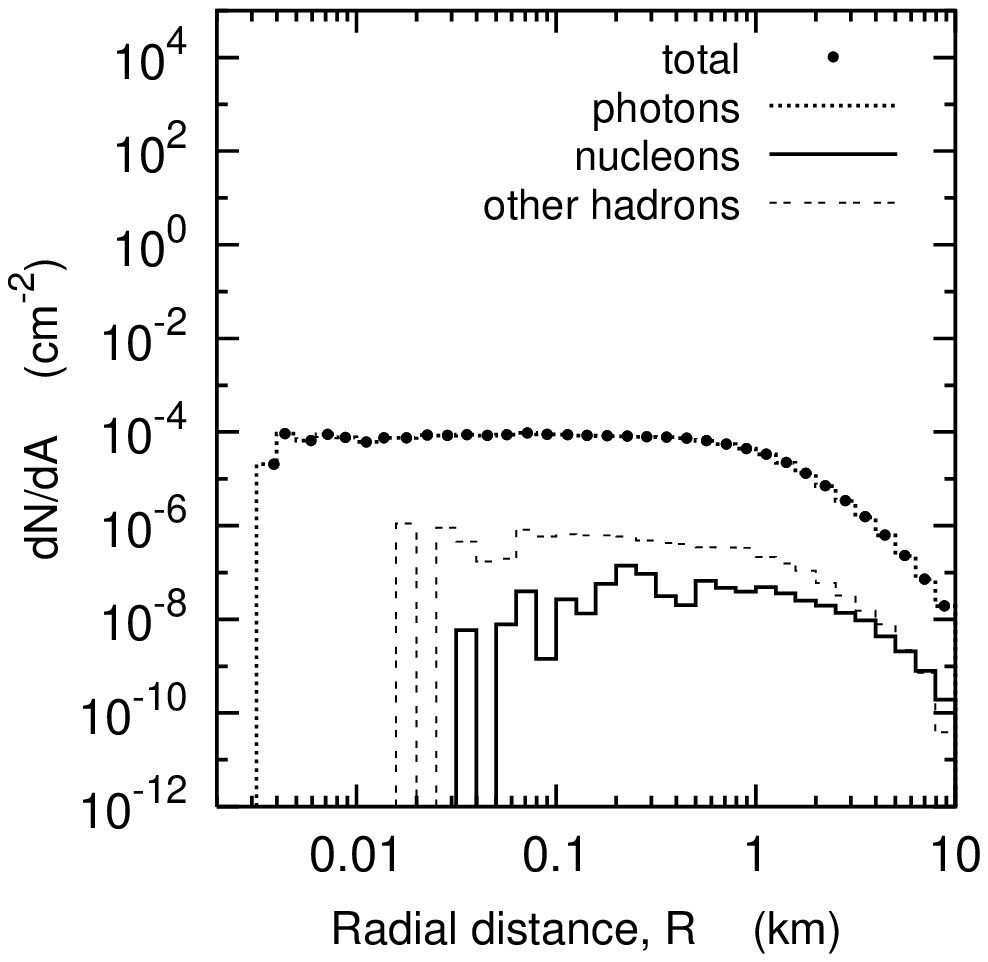, width=100mm}  \\
 \hspace*{20mm}\mbox{(b)}
 \caption{
         }
 \label{dNdA10GeV}
\end{figure}

\begin{figure}[htb]
 \vspace*{5cm}
 \hspace*{-10mm}\epsfig{file=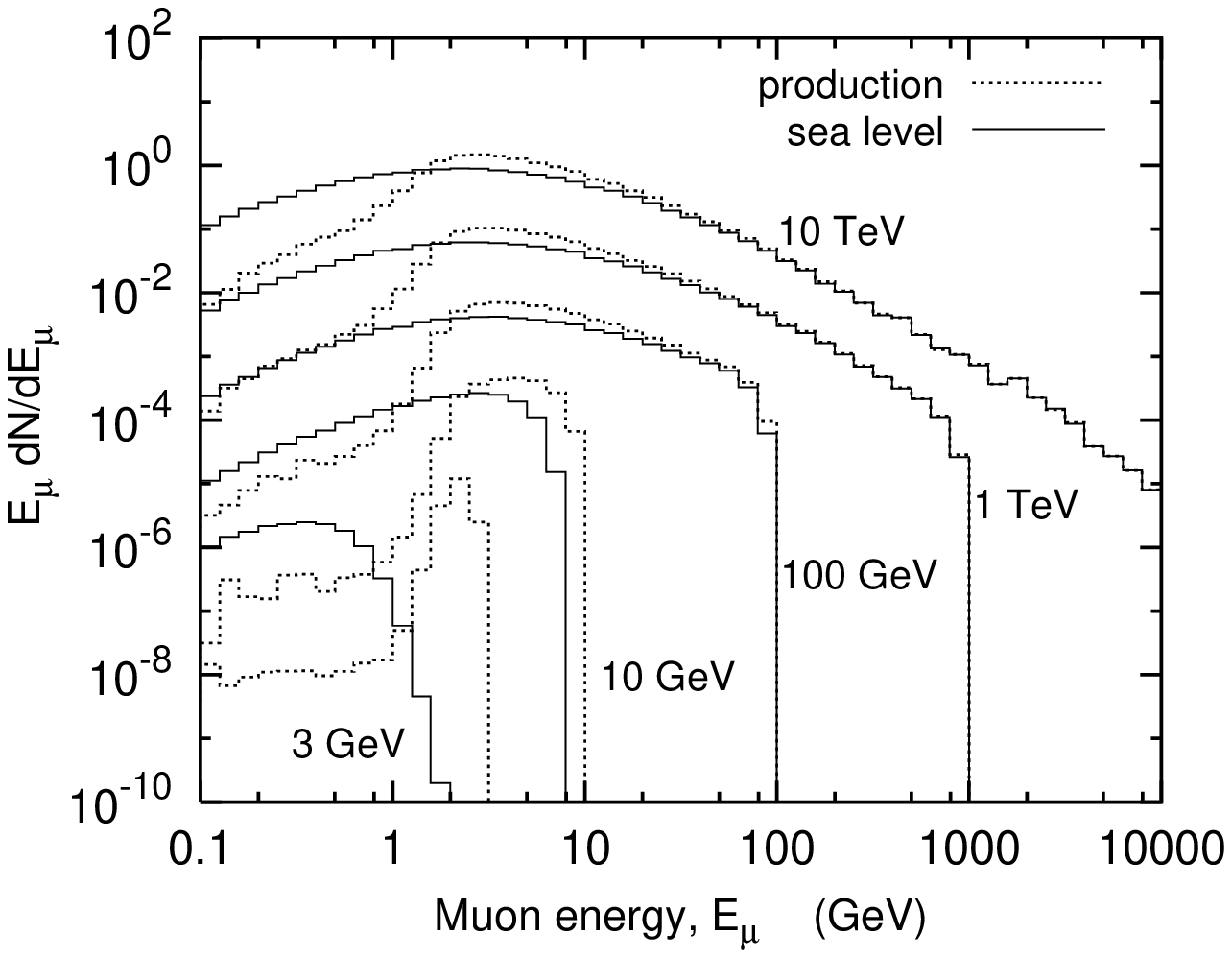,width=13cm}
 \caption{
         }
 \label{dNdEkin}
\end{figure}

\begin{figure}[htb]
 \vspace*{3cm}
 \hspace*{10mm}\epsfig{file=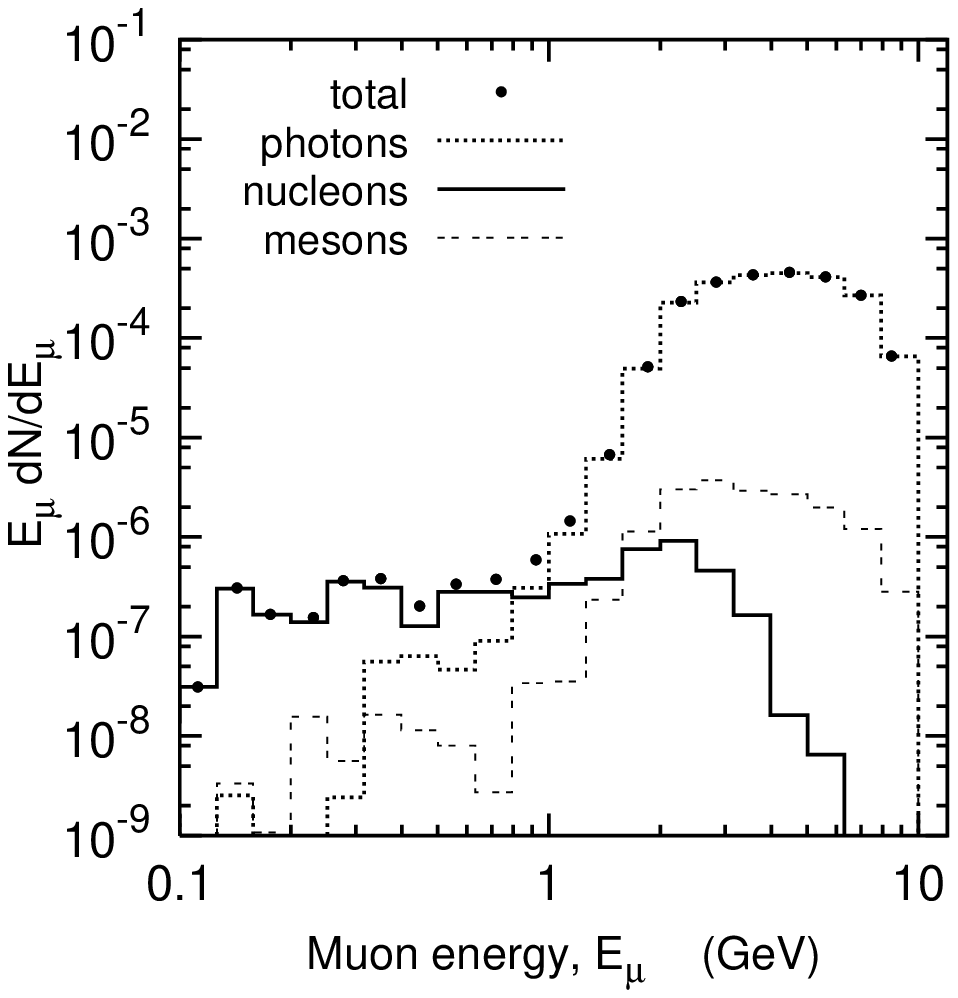,width=100mm}\\
 \hspace*{20mm}\mbox{(a)}\\
 \hspace*{10mm}\epsfig{file=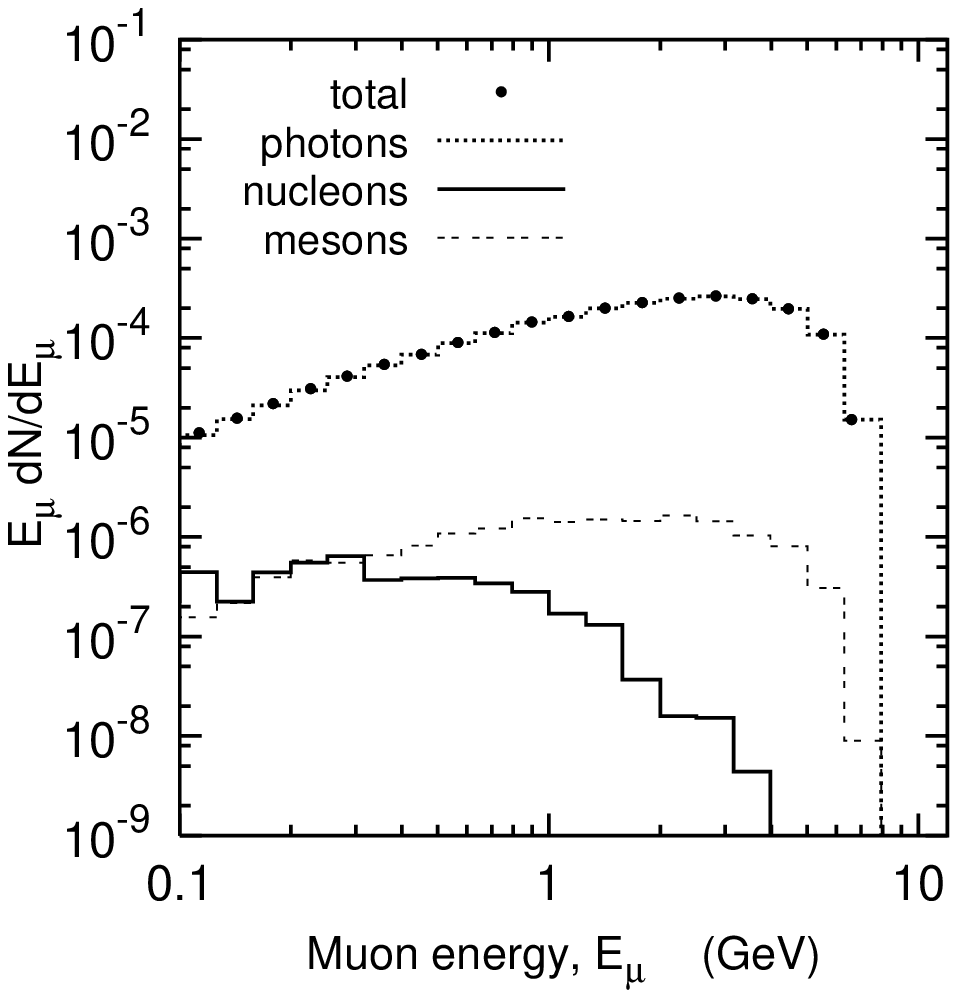,width=100mm}\\
 \hspace*{20mm}\mbox{(b)}
 \caption{
         }
 \label{dNdEkin10GeV}
\end{figure}

\begin{enumerate}
\item[\ref{mu-capdep}.]
    Dependence of the negative muon flux produced by hadronic primaries
    on the depth in the atmosphere 
    shown for different intervals of the muon momentum. {\sc FLUKA} results 
    (solid lines) are compared to data obtained by 
    the CAPRICE experiment~\protect\cite{Boe00a} (points). The comparison 
    is shown for different intervals in muon momentum. The curves and data 
    points for the first five momentum intervals were shifted by constant 
    factors as indicated.
\item[\ref{mu-capspc25&ratio}.]
    a) Energy spectra of positive and negative muons at 886~g/cm$^2$. 
    {\sc FLUKA} results (histograms) are compared to data 
    obtained by the CAPRICE experiment~\protect\cite{Kre99}. $p_{\rm Lab}$
    is the detected muon momentum.
    b) Muon charge ratio as a function of kinetic energy measured by the CAPRICE
    experiment~\protect\cite{Kre99} (points) and calculated with {\sc FLUKA}
    (solid line).
\item[\ref{dNdTh}.]
    Angular distribution of muons at sea level. $\Theta_{xz}$ is the
    angle of the detected muon projected onto the North-down or $xz$ plane. 
    The distributions are given for different primary photon energies and 
    are normalized to unit area. Effects of the Earth's magnetic field are 
    not included in this figure. Units of angle are in degrees.
\item[\ref{dNdTh-10GTeV}.]
    (a) Angular distribution of muons at sea level for 10~GeV and (b) 10~TeV 
    primary gamma ray energy. The distributions labeled ``total'' are
    the same as shown in Fig.~\protect\ref{dNdTh}. 
    In addition, the contributions
    from muons with kinetic energies below 2~GeV and above 2~GeV are presented.
    Units of angle are in degrees.
\item[\ref{dNdThx-10GeV1TeVx}.]
    Angular distributions of muons at sea level given for different
    intervals in the $x$ coordinate. Results are shown for primary
    gamma ray energies of (a) 10~GeV and (b) 1~TeV. All distributions
    are normalized to unit area. Units of angle are in degrees.
\item[\ref{flxgamma}.]
    (a) The figure shows the following quantities as a 
    function of the primary gamma ray energy: the 
    energy spectrum of primary gamma rays ($d\Phi_\gamma/dE_\gamma$) with 
    a shape given by a differential spectral index $\alpha = 2.41$ and 
    arbitrary normalization, the 
    multiplicity of muons at sea level per primary gamma ray 
    ($N_\mu/N_\gamma$), the result of folding the multiplicity with the 
    primary spectrum ($d\Phi_\mu/dE_\gamma=N_\mu/N_\gamma\times 
    d\Phi_\gamma/dE_\gamma$; thin solid line with circles), 
    the half-widths at half-maximum-height for all muon energies 
    ($\delta\Theta_{xz}$(HWHM)) in degrees, and the result of folding these
    widths with $d\Phi_\mu/dE_\gamma$ (dot-dashed lines with diamond
    points). The calculated values 
    are joined by lines to guide the eye. Note that
    all differential fluxes are multiplied by $E_\gamma$.
    (b) Average half-width at half-maximum-height for the angle projected
    onto the North-down plane as function of the spectral index of the primary 
    spectrum; the ordinate is in units of degrees.  Values for the space 
    angle resolution 
    ($\delta\Theta_{R}$(HWHM)) would be $\sqrt{2}$ times larger.  
\item[\ref{dNdTh-10GeV1TeVmag}.]
    Muon angular distribution at sea level including the effect of the 
    Earth's magnetic field for (a) 10~GeV and (b) 1~TeV 
    primary gamma ray energy. The distributions are shown as a function
    of the muon's angle projected onto the $yz$ (East-down) plane, 
    $\Theta_{yz}$, and for three different geographic locations: (1)
    $42^\circ$N, $86^\circ$W (GRAND), (2) $36^\circ$N, $106^\circ$W (MILAGRO), 
    and (3) $9^\circ$N, $100^\circ$E (maximum). The magnetic deflection for 
    $\Theta_{xz}$ (North-down)
    is minimal and not shown.  In addition, the distribution obtained
    without magnetic field (labeled ``no field'') is given.
    Units of angle are in degrees.
\item[\ref{dNdg}.]
    Distribution in the number of generations in the shower (see text)
    carried by the particles creating the parent-meson.
    {\sc FLUKA} results are shown for
    different energies of the primary photon.
\item[\ref{dNdh}.]
    Distribution of the muon production heights. Results are given for
    primary  photon energies of $3, 10, 100$, and $1000$~GeV, respectively,
    and are normalized per primary photon.
    The top of the atmosphere is at 80~km in the calculations.
\item[\ref{dNdA}.]
    Radial distribution of the muon production vertices (i.e. meson decay
    vertices, dotted histograms) and of the muons at sea level (solid 
    histograms). Results are given for primary photon energies of $3, 10, 100, 
    10^3,$ and $10^4$~GeV, respectively (from the bottom curve to the top). 
    Values have been normalized per primary photon.
\item[\ref{dNdA10GeV}.]
    Radial distribution of the muon production vertices for 10~GeV primary
    photons at normal incidence. The distributions are given for (a) the muon 
    production vertex and (b) for sea level. In addition 
    to the total distribution, the contributions from different ``parents'' of 
    the decaying meson are given. For example, the histograms labeled 
    ``photons'' show the radial distributions of 
    muons from mesons which were produced by photonuclear interactions.
    Distributions are normalized per primary photon.
\item[\ref{dNdEkin}.]
    Kinetic energy spectra of muons at their production vertices (dotted
    histograms) and at sea level (solid histograms). Values are normalized per
    primary photon.
\item[\ref{dNdEkin10GeV}.]
    Kinetic energy spectra of muons from 10~GeV primary photons at normal
    incidence. The spectra are shown for (a) the production vertex 
    and (b) for sea level. In addition to the total spectra which
    are identical to those shown in Fig.~\protect\ref{dNdEkin}, the 
    contributions from different ``parents'' of the decaying meson are given
    (similar to Fig.~\protect\ref{dNdA10GeV}).
\end{enumerate}
%
%
\newpage
\newpage
\newpage
\newpage
\newpage
\newpage
\newpage
\newpage
\clearpage
\newpage
\newpage
\newpage
\newpage
\newpage
%
%

\begin{thebibliography}{999}
%
\bibitem{Fas00c}
  A.\ Fass\`o and J.\ Poirier,
  Phys.\ Rev.\ D {\bf 63}, 036002 (2001).
\bibitem{MILAGRO} 
  The MILAGRO Collaboration, R. Atkins {\em et al.},
  Nucl.\ Instr.\ Meth.\ in Phys.\ Res.\ A {\bf 449}, 478 (2000).
\bibitem{GRAND} 
  J.\ Poirier {\em et al.}, 
  in {\em Proceedings of the 26th International Cosmic Ray Conference (ICRC)},
  Vol. 5, p.~304 (1999),
  World Wide Web: {\tt http://www.nd.edu/}$\sim${\tt grand}.
\bibitem{Fas97a} 
  A. Fass\`o {\em et al.}, 
  in {\em Proceedings of the 2nd Workshop on Simulating Accelerator Radiation 
  Environments, CERN 1995}, edited by G.R.~Stevenson, CERN Report TIS-RP/97-05, 
  p.~158 (1997).
\bibitem{Fas97b} 
  A.\ Fass\`o {\em et al.}, 
  in {\em Proceedings of the 3rd Workshop on Simulating Accelerator Radiation 
  Environments, KEK 1997}, edited by H.\ Hirayama, KEK Proceedings 97--5,
  p.~32 (1997).
\bibitem{Fas00a} 
  A.\ Fass\`o, A.\ Ferrari, and P.\ R.\ Sala, 
  {\em Electron-photon Transport in FLUKA: Status},
  to appear in {\em Proceedings of the International Conference on
  Advanced Monte Carlo for Radiation Physics, Particle Transport Simulation 
  and Applications, Monte Carlo 2000}, Lisbon, Portugal, 2000.
\bibitem{Fas00b} 
  A.\ Fass\`o {\em et al.}, 
  {\em FLUKA: Status and Perspectives for Hadronic Applications},
  to appear in {\em Proceedings of the International Conference on
  Advanced Monte Carlo for Radiation Physics, Particle Transport Simulation 
  and Applications, Monte Carlo 2000}, Lisbon, Portugal, 2000.
\bibitem{Pat95} 
  V.\ Patera {\em et al.}, 
  Nucl.\ Instr.\ Meth.\ in Phys.\ Res.\ A {\bf 356}, 514 (1995).
\bibitem{Fer97a} 
  A.\ Ferrari, T.\ Rancati, and P.\ R.\ Sala,
  in {\em Proceedings of the 3rd Workshop on Simulating Accelerator Radiation 
  Environments, KEK 1997}, edited by H.\ Hirayama (KEK Proceedings 97--5, 1997),
  p.~165.
\bibitem{Roe98a}
  S.\ Roesler, W.\ Heinrich, and H.\ Schraube, 
  Rad.\ Res.\ {\bf 149}, 87 (1998).
\bibitem{Bat00} 
  G.\ Battistoni {\em et al.},
  Astropart.\ Phys.\ {\bf 12}, 315 (2000).
\bibitem{Cap94} 
  A.\ Capella {\em et al.}, 
  Phys. Rep. {\bf 236}, 225 (1994).
\bibitem{Bau78}
  T.\ H.\ Bauer, R.\ D.\ Spital,  and D.\ R.\ Yennie,
  Rev.\ Mod.\ Phys.\ {\bf 50}, 261 (1978).
\bibitem{Eng95}
  R.\ Engel,
  Z.\ Phys.\ C {\bf 66}, 203 (1995).
\bibitem{Roe98b}
  S.\ Roesler, R.\ Engel, and J.\ Ranft,
  Phys.\ Rev.\ D {\bf 57}, 2889 (1998).
\bibitem{Boe00a}
  M.\ Boezio {\em et al.},
  Phys.\ Rev.\ D {\bf 62}, 032007 (2000).
\bibitem{Roe00} 
  S.\ Roesler, W.\ Heinrich, and H.\ Schraube, 
  {\em Monte Carlo simulation of the radiation field at aircraft altitudes}, 
  in preparation.
\bibitem{Kre99}
  J.\ Kremer {\em et al.},
  Phys.\ Rev.\ Lett.\ {\bf 83}, 4241 (1999).
\bibitem{Har99} 
  R.\ C.\ Hartman {\em et al.},  
  Astrophys.\ J.\ Suppl.\ Ser.\ {\bf 123}, 79 (1999).
\bibitem{IGRF2000} 
  International Association of Geomagnetism and Aeronomy (IAGA),
  Division V, Working Group 8, 
  {\em International Geomagnetic Reference Field - Epoch 2000, Revision 
  of the IGRF for 2000 - 2005},
  World Wide Web: {\tt http://www.ngdc.noaa.gov/IAGA/wg8/wg8.html}
%
\end{thebibliography}
\end{document}